\documentclass[12pt]{article}
\usepackage{epsfig}
\begin{document}
\begin{center}
{\Large \textbf{Field distribution analysis in deflecting structures.}}\\
\vspace*{0.8cm}
V.V. Paramonov \\
Institute for Nuclear Research, 117312, Moscow, Russia
\end{center}
\begin{abstract} 
Deflecting structures  are used now manly for bunch rotation in emittance exchange concepts, bunch diagnostics 
and to increase the luminosity. The bunch rotation is a transformation of a particles distribution in 
the six dimensional phase space. Together with the expected transformations, deflecting structures introduce 
distortions due to particularities - aberrations - in the deflecting field distribution. 
The distributions of deflecting fields are considered with respect to non linear additions, which 
provide emittance deteriorations during a transformation. The deflecting field is treated as combination of 
hybrid waves $HE_1$ and $HM_1$. The criteria for 
 selection and formation of deflecting structures with minimized level of aberrations are 
formulated and applied to known structures. Results of the study are confirmed by comparison with 
results of numerical simulations.  
\end{abstract}
\newpage
\tableofcontents
\newpage
\section{Introduction} 
 In particle accelerators Deflecting Structures (DS) - periodical structures with transverse 
components of the electromagnetic field at the axis - were introduced for charged particle 
deflection and separation. A bunch of charged particles crosses a DS synchronously with the maximal 
deflecting field $Ed$, corresponding to a phase $\phi=0$ in the structure, and particles get an 
increment in the transverse momentum $p_t$. It allows both to deflect particles from the axis and to separate 
particles with different charge and momentum in space, see, for example, 
\cite{slac_sep}.\\ 
In the modern facilities with short and bright bunches DS found other applications, 
such as short bunch rotation for special diagnostics, emittance exchange experiments and luminosity 
enhancement. All these applications are related to the transformation of particle distributions in 
the six dimensional phase space. For these applications the DS operates in another mode - the center of 
the bunch crosses the DS at zero value of $E_d$, corresponding to $\phi=\frac{\pi}{2}$, see, for example \cite{bu_rot}.\\
The applications for particle distribution transformations provide additional specific 
requirements. Usual RF parameters, like RF efficiency, field rise time, total deflecting voltage $V_d$, describe 
general parameters - achievable resolution of measurements, possibility for single bunch measurements 
in the bunch train and the price for this with respect to the RF system.\\
But, together with the expected transformations, DS provide distortions due to particularities in the 
deflecting field distribution. The tool for distribution transformation should provide as minimal 
as possible intrinsic distortions.\\ 
In a complicated DS geometry the distribution of the electromagnetic field components can be obtained with 
good precision only in numerical simulations. And the distortions of particle distributions 
can be estimated quantitatively also only in numerical simulations of the beam dynamics.\\
As it is known from theory, a system with linear spatial distribution of the field components 
doesn't change the bunch emittance. 
To avoid multiple coupled simulations of field distributions and particle dynamics for different 
possible DS solutions and bunch parameters, we investigate first the different DS options for 
criteria of field linearity, i.e. minimal deviations from a linear field distribution.
\section{Field description}
In the periodical structure the field distribution for each $j$-th component $E_{j}(\vartheta,r,z)$ 
can be represented in complex form and satisfies the Floquet theorem:
\begin{eqnarray}
E_{j}(\vartheta,r,z)=\widehat{E_j(\vartheta,r,z)} e^{i \psi_j (\vartheta,r,z)}
 \quad  -d/2 \leq z < d/2,\\
\nonumber 
E_{j}(\vartheta,r,z+nd)=\widehat{E_j(\vartheta,r,z)} e^{i (\psi_j (\vartheta,r,z)-n\Theta_0)},\quad 0 \leq \Theta_0 \leq \pi, 
\label{2.1e}
\end{eqnarray}
where $\widehat{E_j(\vartheta,r,z)}$ and $\psi_j (\vartheta,r,z)$ are the amplitude and the phase distributions
of the field components at the period, correspondingly, $d$ is the period length, $\Theta_0$ is the 
phase advance per period, and $n$ is the period number.\\
If the structure has a planes of mirror symmetry (the mostly realized case in practice), $\widehat{E_j(\vartheta,r,z)}$ 
is all time an \textbf{even} function with respect to the mirror plane and $\psi_j (\vartheta,r,z)$ is an 
\textbf{odd} function with a possible total shift at $\pi/2$ or $\pi$.\\
In the beam aperture of a slow wave structure the field components can be expanded, see for example
 \cite{lapost}, into a Fourier series over the spatial harmonics:
\begin{equation}
E_{j}(\vartheta,r,z)=\sum^{p \rightarrow + \infty}_{p \rightarrow - \infty} a_{jp}(\vartheta,r) e^{-i k_{zp} z},\quad
k_{zp} =\frac{\Theta_0 + 2 p \pi}{d}, \quad p=0, \pm 1, \pm 2, ..., \pm \infty
\label{2.2e}
\end{equation}
with
\begin{equation}
a_{jp}(\vartheta,r) = \frac{1}{d} \int_{-\frac{d}{2}}^{\frac{d}{2}} E_{j}(\vartheta,r,z)e ^{i k_{zp} z}dz.
\label{2.3e}
\end{equation}
where $a_{jp}(\vartheta,r)$ and $k_{zp}$ are the amplitude and wave number of the $p$-th spatial harmonic,
respectively. Taking into account parity properties for the amplitude and the phase in 
(\ref{2.1e}), the functions $a_{jp}(\vartheta,r)$ in (\ref{2.2e}) are either real or imaginary.\\ 
To avoid the additional introduction of symbols, below we will use and assume the symbols for spatial harmonics in two 
options. If the symbol $a_{jp}(\vartheta,r)$ is used, it means the function with respect $\vartheta,r$, 
including the constant amplitude coefficient. And the symbol $a_{jp}$, without dependence on the coordinates, 
means the constant amplitude coefficient for respective spatial harmonic.\\
For the slow wave system the spatial harmonics in the field representation (\ref{2.2e}) are necessary 
for the boundary conditions at the aperture radius $r=a$. In the aperture volume both total field and 
each spatial harmonic should satisfy to the Maxwell equations. 
\subsection{Relations between field components}
For the complete description of the field distribution two independent variables are required. 
For structures periodical in $z$ it is common practice to consider the
longitudinal components $E_z$ and $H_z$ as such independent variables.\\
From Maxwell equations 
\begin{equation}
rot \vec H = \frac{\partial \vec D}{\partial t}, \quad rot \vec E = -\frac{\partial \vec B}{\partial t},
\quad div \vec B=0,\quad \quad div \vec D=0,\quad \vec D= \epsilon_0 \epsilon \vec E,\quad \vec B= \mu_0 \mu \vec H,
\label{2.1.1e}
\end{equation}
and assuming a $\sim e^{i \omega t}$ time dependence in vacuum, i.e.  $\epsilon =\mu =1$, one can get in 
cylindrical coordinates $r,\vartheta,z$: 
\begin{eqnarray}
\frac{\partial ^2 E_r}{\partial z^2} +k^2 E_r = \frac{\partial ^2 E_z}{\partial r \partial z} - 
\frac{i k n Z_0}{r} H_z,\\
\nonumber
\frac{\partial ^2 E_{\vartheta}}{\partial z^2} +k^2 E_{\vartheta} = -\frac{n}{r} \frac{ \partial E_z}{ \partial z} + 
 i k Z_0 \frac{\partial H_z}{\partial r},\\
\nonumber
Z_0 \frac{\partial ^2 H_r}{\partial z^2} +k^2 Z_0 H_r = -\frac{i k n}{r} E_z +Z_0 \frac{\partial ^2 H_z}{\partial r \partial z},\\
\nonumber
Z_0 \frac{\partial ^2 H_{\vartheta}}{\partial z^2} +k^2 Z_0 H_{\vartheta} = - ik \frac{ \partial E_z}{ \partial r} + 
 \frac{n Z_0}{r} \frac{\partial H_z}{\partial z},
\label{2.1.2e}
\end{eqnarray}
where 
\begin{equation}
k=\frac{\omega}{c} = \omega \sqrt{\epsilon_0 \mu_0}, \quad Z_0= \sqrt{\frac{\mu_0}{\epsilon_0}},
\quad 
(E_z, E_r, H_{\vartheta}) \sim cos(n \vartheta),\quad (H_z, H_r, E_{\vartheta}) \sim sin(n \vartheta).
\label{2.1.3e}
\end{equation}
Also in the cylindrical coordinate system one can get the Bessel equation for the spatial harmonics $f(r)$ of the longitudinal 
field components $E_z(r)$ or $H_z(r)$ from Maxwell equations:
\begin{equation}
\frac{d^2 f(r)}{dr^2}+\frac{1}{r} \frac{df(r)}{dr}+(k^2_{sp}-\frac{n^2}{r^2})f(r) =0, \quad k^2_{sp}=k^2-k^2_{zp},
\label{2.1.5e}
\end{equation}
with the finite solutions at $r=0$:
\begin{equation}
 f(r) = J_n(k_{sp} r ), \quad k^2_{sp} > 0 \quad or \quad f(r) = I_n(-i k_{sp} r ), \quad k^2_{sp} < 0, 
\label{2.1.6e}
\end{equation}
where $J_n(x)$ and $I_n(x)$ are Bessel functions of the first order.\\
In representation (\ref{2.2e}) all spatial harmonics have different relative phase velocities $\beta_{p}$ :
\begin{equation}
 \beta_{p} = \frac{v_p}{c}=\frac{\omega}{c k_{zp}}=\frac{k}{k_{zp}}=\frac{kd}{\Theta_0 + 2 p \pi}. 
\label{2.1.7e}
\end{equation}
The period length $d$ is normally chosen for the synchronous interaction of the particle, moving with 
velocity $v = \beta c, \quad \beta \leq 1$, with a specified, usually the main spatial harmonic 
($p=0$ in (\ref{2.2e})) and 
\begin{equation}
 \beta = \beta_0 \Rightarrow \beta =\frac{kd}{\Theta_0}, \quad d=\frac{\Theta_0 \beta}{k} = 
\frac{\Theta_0 \beta \lambda}{2 \pi}, 
\label{2.1.8e}
\end{equation}
where $\lambda$ is the operating wavelength.
From (\ref{2.1.8e}, \ref{2.1.7e}, \ref{2.2e}) follows $|\beta_p| < 1 $ for all $p \neq 0$, resulting 
always in $k^2_{sp} = k^2(1-\frac{1}{\beta_p^2}) \leq 0$, see (\ref{2.1.6e}). The modified Bessel functions 
$I_n(x)$ describe the radial dependences of the spatial harmonics in the representation of the field components 
(\ref{2.2e}).
\subsection{Properties of Bessel functions}
The general expansion of $I_n(x)$ in a power series in $x$ is: 
\begin{equation}
 I_n(x)= (\frac{x}{2})^n \sum^{ \infty}_{j=0}
\frac{(\frac{x}{2})^{2j}}{j!\Gamma(n+j+1)}, \quad I_n^{(1)}(x)=\frac{dI_n(x)}{dx}= I_{n-1}(x) -
\frac{n}{x} I_n(x), 
\label{2.2.1e}
\end{equation}
where $\Gamma(n+j+1)$ is the Gamma function for integer arguments.\\
For large arguments $x \gg 1$ the functions $I_n(x)$ rises exponentially with increasing argument 
\begin{equation}
 I_n(x) \approx  \frac{e^x}{\sqrt{2 \pi x}}.
\label{2.2.2e}
\end{equation}  
For the approximate estimation of the field components near the axis (x= $ k^*_{sp}r \ll 1$) and for the main 
spatial harmonic with $\beta_0 \rightarrow 1$ we can obtain from (\ref{2.2.1e}):
\begin{equation}
 I_n(x) \approx  \frac{x^n}{2^n (n-1)!}+ \frac{x^{n+2}}{2^{n+2} n!}+...,\quad I_0(x) \approx 1+\frac{x^2}{4}+..., 
\quad I_1(x) \approx \frac{x}{2} +\frac{x^3}{8}+...,
\label{2.2.3e}
\end{equation}  
\subsection{Field components of dipole mode}  
For the longitudinal components $E_z$ and $H_z$ of the dipole mode $n=1$ representation (\ref{2.2e}) can be 
rewritten as:
\begin{eqnarray}
E_{z}(\vartheta,r,z)=cos(\vartheta)\sum^{p \rightarrow + \infty}_{p \rightarrow - \infty} 
e_{zp}I_1( k^*_{sp} r ) e^{-i k_{zp} z}, \quad k^*_{sp}=-i \sqrt{k^2_{sp}},\\
\nonumber
Z_0 H_{z}(\vartheta,r,z)=sin(\vartheta)\sum^{p \rightarrow + \infty}_{p \rightarrow - \infty} 
Z_0 h_{zp}I_1(k^*_{sp} r) e^{-i k_{zp} z}.
\label{2.3.1e}
\end{eqnarray} 
Relations for coefficients and radial dependencies for spatial harmonics in the other field components 
can be expressed from (5) as:
\begin{eqnarray}
k^2_{sp}  (r)e_{rp}(r)=-i k_{zp} k^*_{sp} I_1^{(1)}(k^*_{sp} r) e_{zp} - \frac{ik }{r} I_1(k^*_{sp} r) Z_0 h_{zp},\\
\nonumber
k^2_{sp}   e_{\vartheta p}(r)=\frac{i k_{zp} }{r} I_1(k^*_{sp} r)e_{zp} + ik k^*_{sp} I_1^{(1)}(k^*_{sp} r) Z_0 h_{zp},\\
\nonumber
 k^2_{sp}  Z_0 h_{rp}(r)=- \frac{ik}{r} I_1(k^*_{sp} r) e_{zp} -i k_{zp} k^*_{sp} I_1^{(1)}(k^*_{sp} r) Z_0 h_{zp},\\
\nonumber
k^2_{sp}   Z_0 h_{\vartheta p}(r) =- ik k^*_{sp} I_1^{(1)}(k^*_{sp} r)e_{zp}- \frac{i k_{zp} }{r} I_1(k^*_{sp} r) Z_0 h_{zp}.
\label{2.3.2e}
\end{eqnarray}
\subsection{Basis problem. Hybrid waves $HE$, $HM$}
\begin{figure}[htb]
\centering
\epsfig{file=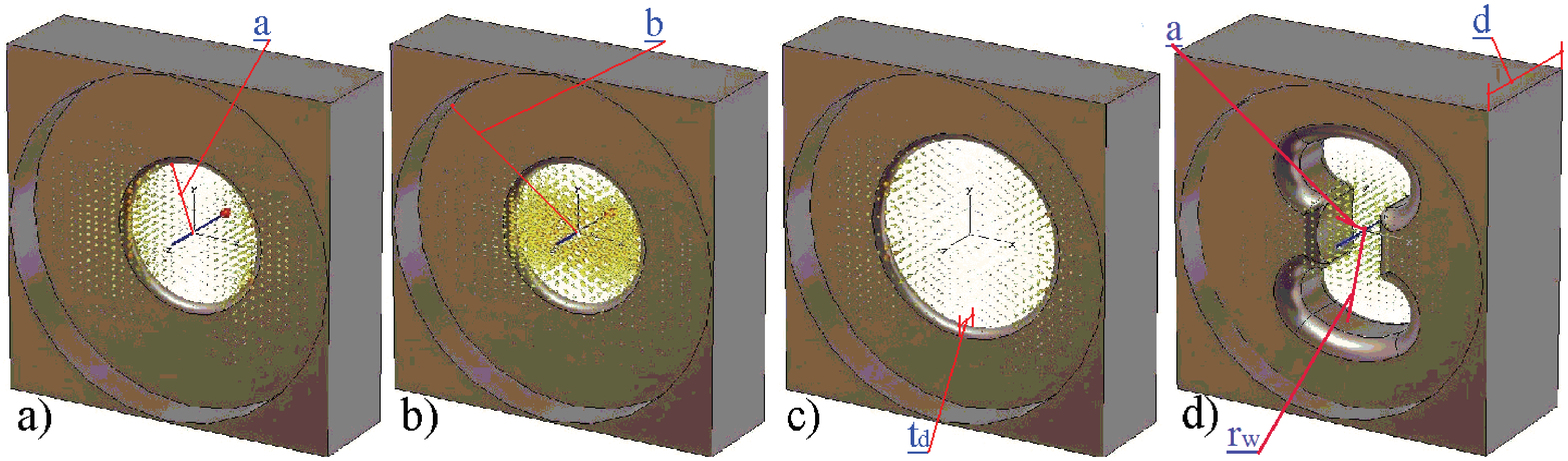, width =160.0mm}
\begin{center}
 Figure 1: Field distributions for waves with $\Theta_0=\frac{2\pi}{3}$ with deflecting effect for different 
structures and modes. a) - with $\frac{a}{\lambda}=0.23 $ at the first DLW passband, $\beta_g <0$, 
b) - with $\frac{a}{\lambda}=0.23 $ at the second DLW passband, $\beta_g >0$,
c) - with $\frac{a}{\lambda}=0.30 $ at the first DLW passband, $\beta_g >0$, 
d) - with $\frac{a}{\lambda}=0.067 $ at the first passband for a decoupled TE-type structure \cite{te_defl},
$\beta_g >0$.
\end{center}
\label{1f}
\end{figure}  
The representation (5) is the well known description 
of an arbitrary field in terms of transversal electric $TE$ and transversal magnetic $TM$ waves. 
In (\ref{2.3.1e}) it is detailed for a slow wave system with spatial harmonics. In a mathematical sense 
it is a basis for a field expansion. This basis works always well, except for one point.
For an ultra relativistic particle $\beta=1$ the synchronous main spatial harmonic $p=0$ 
has $k_{z0}=k$, (2,9) and $k_{sp}=0$, (\ref{2.1.6e}) and there is an indefiniteness in the relations 
(15) between the amplitudes of the field components for the main spatial harmonics.\\ 
For  $\beta_0 \rightarrow 1 $ both $TE$ and $TM$ waves degenerate into a simple plane $TEM$ wave with 
transversal components only. Attempts for deflecting field description basing only on $TE-TM$ 
terms are hence not successful.\\
It is just a methodical problem of the description for the real physical objects - the fields with the effect 
of deflection.  In Fig. 1 the field distributions are shown for 
different modes and different structures, but all exhibiting effect of deflection. In Fig. 1a, b, c 
are field distributions in the well known Disk Loaded Waveguide (DLW) shown. The definition of dimensions are: 
$a$ is the aperture radius, $b$ is the cell radius, $t_d$ is the disk thickness,
$\beta_g$ is the group velocity. A TE-type structure, \cite{te_defl}, is shown in Fig. 1d.\\  
Nearly simultaneous investigations in different laboratories, see summarizing papers 
\cite{garo},\cite{hahn},\cite{alex}, resulted in the proposal of another basis for the field 
representation in cylindrical coordinates. This basis was named as 'hybrid' $EH$ in \cite{garo} and 
\cite{alex}, $HM -HE$ in \cite{hahn} and was derived from Herzian vectors as another independent solution 
of the Helmholtz vector equation.\\
Such an introduction of additional elements in the basis of waves was not the first one. $LE$ and $LM$ 
waves were introduced for the $\beta_0=1$ case description in rectangular slow wave systems, 
see the book \cite{slow} and references for earlier papers.\\
Comparison of $TE - TM$ and $HE - HM$ waves is done in \cite{hahn} and expressions 
for field components are reproduced here in the Table 1. 
\begin{table}[htb]   
\begin{center}
\centering{Table 1: Field components for the transverse $TE - TM$ and hybrid $HE - HM$ waves, \cite{hahn}. }
\begin{tabular}{|l|c|c|c|c|c|c|c|}
\hline
                      & $TM$                                          & $TE$                                          & $HM$                                                                          & $HE$                                                                          \\
\hline
                      &                                               &                                               &                                                                               &                                                                               \\  
 $E_r \sim$           & $-i k_z \frac{J^{(1)}_{n}(k_s r)}{k_s^{n-1}}$ & $-i k n \frac{J_{n}(k_s r)}{k_s^{n}r}$        & $i k k_z  \frac{J_{n+1}(k_s r)}{k_s^{n+1}}$                                  & $i(k_z^2 \frac{J_{n+1}(k_s r)}{k_s^{n+1}} +n \frac{J_{n}(k_s r)}{k_s^{n}r}$)   \\
                      &                                               &                                               &                                                                               &                                                                               \\  
\hline
                      &                                               &                                               &                                                                               &                                                                               \\  
 $E_{\vartheta} \sim$ & $-i k_z n \frac{J_{n}(k_s r)}{k_s^{n}r}$      & $i k \frac{J^{(1)}_{n}(k_s r)}{k_s^{n-1}}$    & $i k k_z  \frac{J_{n+1}(k_s r)}{k_s^{n+1}}$                                  & $i(k^2 \frac{J_{n+1}(k_s r)}{k_s^{n+1}} -n \frac{J_{n}(k_s r)}{k_s^{n}r})$     \\
                      &                                               &                                               &                                                                               &                                                                               \\  
\hline
                      &                                               &                                               &                                                                               &                                                                               \\  
 $E_z \sim $          & $ k_s^2 \frac{J_{n}(k_s r)}{k_s^{n}}$        &  0                                            & $ k \frac{J_{n}(k_s r)}{k_s^{n}}$                                             & $ k_z \frac{J_{n}(k_s r)}{k_s^{n}}$                                            \\
                      &                                               &                                               &                                                                               &                                                                               \\  
\hline
                      &                                               &                                               &                                                                               &                                                                               \\  
 $H_r \sim $          & $-i k n \frac{J_{n}(k_s r)}{k_s^{n}r}$        & $-i k_z \frac{J^{(1)}_{n}(k_s r)}{k_s^{n-1}}$ & $-i(k_z^2 \frac{J_{n+1}(k_s r)}{k_s^{n+1}} +n \frac{J_{n}(k_s r)}{k_s^{n}r})$ & $-i k k_z  \frac{J_{n+1}(k_s r)}{k_s^{n+1}}$                                  \\
                      &                                               &                                               &                                                                               &                                                                               \\  
\hline
                      &                                               &                                               &                                                                               &                                                                               \\  
 $H_{\vartheta} \sim$ & $-i k \frac{J^{(1)}_{n}(k_s r)}{k_s^{n-1}}$   & $-i k_z n \frac{J_{n}(k_s r)}{k_s^{n}r}$      & $i(k^2 \frac{J_{n+1}(k_s r)}{k_s^{n+1}} -n \frac{J_{n}(k_s r)}{k_s^{n}r})$    & $i k k_z  \frac{J_{n+1}(k_s r)}{k_s^{n+1}}$                                   \\
                      &                                               &                                               &                                                                               &                                                                               \\     
\hline
                      &                                               &                                               &                                                                               &                                                                               \\  
 $H_z \sim  $         &  0                                            & $ k_s^2 \frac{J_{n}(k_s r)}{k_s^{n}}$        & $ -k_z \frac{J_{n}(k_s r)}{k_s^{n}}$                                          & $ -k \frac{J_{n}(k_s r)}{k_s^{n}}$                                             \\
                      &                                               &                                               &                                                                               &                                                                               \\  
\hline
\end{tabular}
\label{1t}
\end{center}
\end{table} 
The $HE_{n} - HM_{n}$ waves can be treated as hybrid wave with simultaneously existents of all six field 
components. The longitudinal components $E_z$ and $H_z$ are non vanishing for $k_{sp}=0$, but are not 
independent too. For more details of $HE_{n} - HM_{n}$ properties see \cite{hahn} and \cite{garo}.\\
As one can see from the Table 1, taking into account the behavior of the Bessel functions(\ref{2.2.3e}), for a dipole 
mode $n=1$ at the DS axis $r=0$ an $HE_1$ wave has non zero transversal components of the electric field $E_r, E_{\vartheta}$ and 
simultaneously $H_r= H_{\vartheta}=0 $ at $r=0$. Non zero transverse components 
of the magnetic field $H_r, H_{\vartheta}$ at $r=0$ are represented by an $HM_1$ wave only. To describe in a 
real DS the field with simultaneous non zero transverse electric and magnetic components at the axis, 
we need the linear combination 
\begin{equation}
\vec E = A \vec E_{HE}+ B \vec E_{HM}, \quad \vec H = A \vec H_{HE}+ B \vec H_{HM},
\label{2.3.3e}
\end{equation} 
where the coefficients $A, B$ can be defined from the transverse field components $E_r, E_x$ and 
$H_{\vartheta}, H_y$ distributions at the DS axis. 
\subsection{Supporting structure}
The hybrid waves $HE_{n} - HM_{n}$ can not exist without a supporting structure. Differing from $TE_{n} - TM_{n}$
waves, hybrid waves can not exist even in a smooth cylindrical waveguide. In a practical sense $HE_{n} - HM_{n}$
waves are the tool for deflecting effect description and analysis for the main spatial harmonic in the 
beam aperture, where the usual $TE_{n} - TM_{n}$ basis doesn't work for $\beta_0=1$. A preference to describe higher 
spatial harmonics in terms of $HE_{n} - HM_{n}$ is however not evident. It can be done, using the expressions 
for the field components in the Table 1 and (\ref{2.3.3e}), but for all 
harmonics with $p \neq 0$ the more conventional $TE_{n} - TM_{n}$ basis works well.\\
The complete description of DS parameters in terms of $HE - HM$ waves appears not possible, at least it is not 
effective. By using the modern software for the numerical simulation of field distributions and frequency calculations, 
we can estimate all required RF parameters and extract amplitudes of synchronous $HE$ and $HM$ harmonics in the field from the simulated field distribution, basing on 
(\ref{2.3.3e}).\\  
For the DLW structure the solution was obtained in closed form for a small pitch approximation 
$d \ll \lambda, \frac{t_d}{d} \ll 1, \beta=1$ in the first passband of dipole modes. From the approximated 
boundary conditions at $r=a$ both estimation for frequency and for field components (for $r < a$) were 
obtained \cite{garo}:
\begin{eqnarray}
E_z(r,\vartheta)= \frac{E_0}{2} (kr) cos(\vartheta), \quad Z_0 H_z(r,\vartheta)= -\frac{E_0}{2} (kr) sin(\vartheta),\\
\nonumber
E_r(r,\vartheta)= i\frac{E_0}{8} (k^2 a^2 +k^2 r^2) cos(\vartheta), \quad Z_0 H_r(r,\vartheta)= i\frac{E_0}{8} (k^2a^2-k^2r^2 -4) sin(\vartheta),\\
\nonumber
E_{\vartheta}(r,\vartheta)= -i\frac{E_0}{8} (k^2 a^2 -k^2 r^2) sin(\vartheta), \quad Z_0 H_{\vartheta}(r,\vartheta)= i\frac{E_0}{8} (k^2a^2+k^2r^2 -4) cos(\vartheta),
\label{2.5.1e}
\end{eqnarray}
This result is widely used a long time in many papers, see, for example \cite{mont}, and leads to some 
important conclusions.\\
The total flux of traveling RF power $P_{tr}^{tot}$ in a periodical structures is due to the main space harmonic.
Using expressions (17), we obtain:
\begin{equation}
 P_{tr}^{tot} = \frac{\Re}{2} \int_S ([\vec E, \vec H^*],\vec i_z) dS =\frac{\pi k^2 a^4}{32 Z_0}(\frac{k^2 a^2}{3}-1),
\label{2.5.2e}
\end{equation}
For the small aperture radius $ka < \sqrt{3}$ the total power flux $P_{tr}^{tot} < 0, \beta_g < 0$ and 
DLW in the first dipole passband is a backward wave TW structure. For $ka > \sqrt{3}$ the group 
velocity is positive and $\beta_g$ increases with further rise of $a$. The point $ka = \sqrt{3}, \beta_g=0$ 
is known as the inversion point, where $\beta_g$ changes the sign. \\ 
The small pitch approximation
is suitable for the description of a DLW operation in Traveling Wave (TW) mode with very low phase advance 
$\Theta_0 \ll \pi$. For TW mode with $\Theta_0 \geq \pi/2$ or Standing Wave (SW) operation $\Theta_0 = \pi$
the assumption $d \ll \lambda$ is not valid and (17), (\ref{2.5.2e}) are just indications.\\
Considering the DLW deflecting field as the combination of $HE_1$ and $HM_1$ waves, from (\ref{2.3.3e}), 
(17) we can define the ratio for the small pitch approximation as
\begin{equation}
 \frac{B}{A} =  (\frac{Z_0 H_{\vartheta}}{E_r})_{r=0} = \frac{k^2a^2 -4}{k^2a^2}.
\label{2.5.3e}
\end{equation}
\section{Deflecting field representation and analysis}
The Lorenz force acting onto a particle, moving along the $z$ axis with velocity $v=\beta c$ is:
\begin{equation}
\vec F_L = e (\vec E + [\vec v , \vec B]) =e( \vec i_r (E_r -\beta Z_0 H_{\vartheta}) + 
\vec i_{\vartheta} (E_{\vartheta} + \beta Z_0 H_r) + \vec i_z E_z).  
\label{3.1e}
\end{equation}
where $e$ is the electron charge and $\vec i_r, \vec i_{\vartheta}, \vec i_z$ are unit vectors.
The deflecting force $F_d$ and an equivalent deflecting field $E_d$ can be defined through the transverse 
components of the Lorenz force:  
\begin{equation}
\vec F_d = e \vec E_d  =e( \vec i_r (E_r -\beta Z_0 H_{\vartheta}) + 
\vec i_{\vartheta} (E_{\vartheta} + \beta Z_0 H_r)).  
\label{3.2e}
\end{equation}
The deflecting field $E_d$ is the linear combination of the original field components. Regardless to a field 
description in terms of $TM-TE$ or $HM-HE$ waves, each field component can be represented as the sum over spatial 
harmonics, because (\ref{2.2e}) is the sequence of the structure periodicity. 
From linearity, $E_d$ components also can be represented as 
\begin{equation}
E_{dr,d\vartheta}(\vartheta,r,z)=\sum^{p \rightarrow + \infty}_{p \rightarrow - \infty} 
a_{p,dr,d\vartheta}(\vartheta,r) e^{-i k_{zp} z},
\label{3.4e}
\end{equation}
where the amplitudes $a_{p,dr,d\vartheta}(\vartheta,r)$ can be obtained, by using (\ref{3.2e}), from the corresponding 
amplitudes in the expansions of field components.\\
In the cylindrical coordinate system the deflecting field amplitude $\vec E_d$ can be defined, 
at least for the main harmonic $p=0$, \cite{hahn}, from the longitudinal components only, \cite{garo}:
\begin{equation}
\vec F_d = \frac{e}{k_{z0}}  (\frac{1-\beta \beta_0}{1-\beta_0^2} \nabla_t E_z + \frac{\beta-\beta_0}
{1 -\beta_0^2} Z_0[\vec i_z , \nabla_t H_z]), \quad 
\nabla_t =\vec i_r \frac{\partial}{\partial r} +\vec i_{\vartheta} \frac{1}{r}  \frac{\partial}{\partial \vartheta},
\label{3.5e}
\end{equation}
This general expression is valid also for a non synchronous interaction $\beta \neq \beta_0$. 
For synchronous particle motion  $\beta = \beta_0$ relation (\ref{3.5e}) simplifies as:
\begin{equation}
\vec F_d = \frac{e}{k_{z0}} \nabla_t E_z,
\label{3.6e}
\end{equation}
For ultra relativistic particles $(\beta =1)$ the statement (\ref{3.6e}) was derived by Panofsky and 
Wenzel, \cite{pif}, regardless of a classification of the waves as $TM-TE$ or $HM-HE$.
\subsection{Synchronous spatial harmonic}
In the analysis of the main spatial harmonic $p=0$ we have to distinguish two cases - $\beta_0=1$ and 
$\beta_0 <1$.\\
For ultra relativistic particles $\beta = \beta_0=1$. In this case $k_{z0} = k $ in (\ref{2.2e}) and 
$k^2_{s0}=0$ in (\ref{2.1.5e}) - the Bessel equation degenerates into the Laplace equation for $E_z$.
As one can see from Table 1, $HE$ and $HM$ waves are constructed to have non vanishing $E_z$ and 
$H_z$ components for $k^2_{s0} \rightarrow 0$. Expressions for $HE$ and $HM$ field components in 
the case $\beta_0=1, k^2_{s0}=0$ also are considered in \cite{hahn} and reproduced here in the 
Table 2.
\begin{table}[htb]   
\begin{center}
\centering{Table 2: Field components for hybrid $HE - HM$ waves with $\beta_0 =1$, \cite{hahn}.}
\begin{tabular}{|l|c|c|c|c|c|c|c|}
\hline
                      & $HM$                                                                     & $HE$                                                                  & $\vartheta$          \\
\hline
                      &                                                                          &                                                                       &                      \\ 
 $E_r \sim$           & $i \frac{k^2 r^{n+1}}{2^{n+1} (n+1)!}$                                   & $i (\frac{k^2 r^{n+1}}{2^{n+1}(n+1)!} + \frac{ r^{n-1}}{2^n (n-1)!})$ & $cos(n \vartheta)$   \\
                      &                                                                          &                                                                       &                      \\ 
\hline
                      &                                                                          &                                                                       &                      \\ 
 $E_{\vartheta} \sim$ & $i \frac{k^2 r^{n+1}}{2^{n+1} (n+1)!}$                                   & $i (\frac{k^2 r^{n+1}}{2^{n+1}(n+1)!} - \frac{ r^{n-1}}{2^n (n-1)!})$ & $sin(n \vartheta)$   \\
                      &                                                                          &                                                                       &                      \\ 
\hline
                      &                                                                          &                                                                       &                      \\ 
 $E_z \sim $          & $ \frac{k r^n}{2^n n!}$                                                  & $ \frac{k r^n}{2^n n!}$                                               & $cos(n \vartheta)$   \\
                      &                                                                          &                                                                       &                      \\ 
\hline
                      &                                                                          &                                                                       &                      \\ 
 $H_r \sim $          & $-i (\frac{k^2 r^{n+1}}{2^{n+1} (n+1)!} + \frac{r^{n-1}}{2^n (n-1)!})$   & $-i \frac{k^2 r^{n+1}}{2^{n+1}(n+1)!}$                                & $sin(n \vartheta)$   \\
                      &                                                                          &                                                                       &                      \\ 
\hline
                      &                                                                          &                                                                       &                      \\ 
 $H_{\vartheta} \sim$ & $i (\frac{k^2 r^{n+1}}{2^{n+1} (n+1)!} - \frac{r^{n-1}}{2^n (n-1)!})$  & $i \frac{k^2 r^{n+1}}{2^{n+1}(n+1)!}$                                   & $cos(n \vartheta)$   \\
                      &                                                                          &                                                                       &                      \\ 
\hline
                      &                                                                          &                                                                       &                      \\ 
 $H_z \sim  $         & $- \frac{k r^n}{2^n n!}$                                                 & $- \frac{k r^n}{2^n n!}$                                              & $sin(n \vartheta)$   \\
                      &                                                                          &                                                                       &                      \\ 
\hline
\end{tabular}
\label{2t}
\end{center}
\end{table} 
By using (\ref{3.6e}) and $E_z= \frac{E_0 k r}{2}$ for a dipole mode according to Table 2, one can 
directly get:
\begin{equation}
\vec F_d = \frac{e E_0}{2} (\vec i_r cos(\vartheta)- \vec i_{\vartheta} sin(\vartheta)) =\frac{e E_0}{2} \vec i_x.
\label{3.1.1e}
\end{equation}
For $\beta_0=1$ the deflecting force from the synchronous harmonic, both for $HE_1$ and $HM_1$ waves, is constant, 
both in value and in direction, in all points inside the DS aperture. Due to the linearity in (\ref{3.6e}), this statement is 
also valid for the total field (\ref{2.3.3e}). The deflecting force from the synchronous harmonic 
is free from aberrations.\\
For the lower phase velocity, $\beta_0 <1$, the longitudinal component $E_z \sim  k \frac{J_{1}(k_{s0} r)}{k_{s0} }$ for the $HM_1$
wave and $E_z \sim k_{z0} \frac{J_{1}(k_{s0} r)}{k_{s0}}$ for the $HE_1$ wave, see Table 1. Considering the
combination $A\cdot HE_1 + B\cdot HM_1$, applying (\ref{3.6e}) and taking into account $k^2_{s0} < 0$, we get:
\begin{equation}
\vec F_d = \frac{e(A k_{z0} +B k) k_{s0}^* }{k_{z0} k_{s0}} (\vec i_r I_{1}^1(k_{s0}^* r) cos(\vartheta)- \vec i_{\vartheta} \frac {I_{1}(k_{s0}^* r)} {k_{s0}^* r} sin(\vartheta)),
\label{3.1.2e}
\end{equation}
Using (\ref{2.2.1e}) and the approximated expansion (\ref{2.2.3e}), in cylindrical coordinates we have:
\begin{equation}
\vec F_d \approx \frac{e (A k_{z0} +B k)k_{s0}^* }{2 k_{z0} k_{s0}} (\vec i_r (1 +\frac{(k_{s0}^* r)^2}{4}) cos(\vartheta)- 
 \vec i_{\vartheta} (1 + \frac{(k_{s0}^* r)^2} {4}) sin(\vartheta)),
\label{3.1.3e}
\end{equation}
We see in the deflecting force non linear additions $ \sim (k_{s0}^* r)^2$. Transferring (\ref{3.1.3e}) 
into Cartesian coordinates, we have:
\begin{equation}
\vec F_d \approx \vec i_x F_x = \vec i_x \frac{e (A k_{z0} +B k) }{2 k_{z0}} (1 +\frac{(k_{s0}^* x)^2+(k_{s0}^* y)^2}{4}), \quad F_y \approx 0,
\label{3.1.4e}
\end{equation}
The conclusion $F_y \approx 0$ in (\ref{3.1.4e}) is due to the expansion limitation in (\ref{2.2.3e}) 
for each Bessel function with two first terms and with additional higher terms $F_y \sim (k_{s0}^*)^4 (x^3 y+xy^3)$. \\
For not relativistic case the deflecting force is not free from aberrations - there are non linear 
additions even in the force from the synchronous harmonic. These inevitable additions are proportional 
to the constant term in the force and vanish as $\frac{(1 - \beta^2)}{\beta^2} = \frac{1}{\gamma^2 \beta^2}$
 for $\beta \rightarrow 1$, where $\gamma$ is the Lorenz factor.
The non linear additions in deflecting field $x$ component are always even functions with respect to the
planes $x=0, y=0$ while the $y$ field component are odd functions. \\
Based on the conclusions (\ref{3.1.1e}) and (\ref{3.1.4e}) that the deflecting force is directed in the $x$ 
direction, for simplification of the further analysis of numerical results for different DS's, we will 
assume equivalent deflecting field at the axis as:
\begin{equation}
E_d = E_x - Z_0 \beta H_y, 
\label{3.1.5e}
\end{equation}
neglecting a possible $y$ component. Transversal $y$ component in the deflecting field can be due to non linear 
additions only. Moreover, in conversion from cylindrical to Cartesian coordinate systems at the DS axis 
$r=0$ for $\vartheta=0, E_r=E_x, H_{\vartheta}=H_y$.\\
From (\ref{3.1.5e}) we can define coefficients $A, B$ in the total field representation, 
because $E_x =A E_x(HE_1)$ from the $HE_1$ component and $H_y = H_y (HM_1)$ from the $HM_1$ component in (\ref{2.3.3e}).    
\subsection{Higher spatial harmonics for dipole mode}
The higher spatial harmonics $p \neq 0$ in the $E_d$ representation (\ref{3.4e}) always have a low phase 
velocity $|\beta_p| < 1$ and $|k_{sp}| \sim |k_{zp}| > k$. Relation 
(\ref{3.6e}) for higher harmonics is not proven in \cite{hahn} and we will use general approach. 
For transformation reduction, let us consider for beginning spatial harmonics in $HE_1$ wave only. According 
to Table 1, the $p$-th spatial harmonic in the $E_z$ component is:
\begin{equation}
e_{zp}(r) \sim k_{zp} \frac{J_{1}(k_{sp} r)}{k_{sp}},\quad or \quad e_{zp}(r) = C_p I_{1}(k_{sp}^* r), 
\quad C_p=const_1  \cdot \frac{k_{zp}}{k_{sp}}.  
\label{3.2.1e}
\end{equation}
Because $E_z$ and $H_z$ components are not independent in hybrid waves, the respective harmonic in the $H_z$ 
representation is: 
\begin{equation}
h_{zp}(r)  = D_p I_{1}(k_{sp}^* r), 
\quad D_p= - \frac{ C_p k}{ k_{zp}}.  
\label{3.2.2e}
\end{equation} 
To define the amplitudes of the spatial harmonics in the transverse field components, we use relations (15), 
which are a direct consequence of the Maxwell equations. Following (\ref{3.2e}), the amplitudes of 
$p$-th spatial harmonics in the $E_d$ components, assuming $\beta=1$, are: 
\begin{eqnarray}
e_{drp}(r) = e_{rp}(r) - Z_o h_{\vartheta p}(r) \approx \frac{i( k_{zp}-k)}{2 k^*_{sp}}(C_p -D_p)(1+\frac{(k^*_{sp} r)^2}{4})e_{zp},\\
\nonumber
e_{d \vartheta p}(r)= e_{\vartheta p}(r) + Z_0 h_{rp}(r) \approx \frac{-i( k_{zp}-k)}{2 k^*_{sp}} (C_p -D_p)(1+\frac{(k^*_{sp} r)^2}{4}) e_{zp}.
\label{3.2.4e}
\end{eqnarray}
where the approximated expansion (\ref{2.2.3e}) is applied. From (\ref{3.2.4e}) we get in Cartesian 
coordinates for the spatial harmonic in the deflecting field due to the $HE_1$ wave:
\begin{equation}
e_{dxp}(x,y) \approx \frac{i( k_{zp}-k)}{2 k^*_{sp}}(C_p -D_p)(1+\frac{(k^*_{sp} x)^2+(k^*_{sp} y)^2}{4})e_{zp},\quad e_{dyp}(x,y) \approx 0.
\label{3.2.5e}
\end{equation}
A similar expression, but assuming $C_p=const_2  \cdot \frac{k}{k_{sp}}, D_p= - \frac{ C_p k_{zp}}{ k}$ 
can be obtained for the spatial harmonic in the deflecting field due to the $HM_1$ wave.\\
Considering the deflecting field as combination $A\cdot HE_1 + B\cdot HM_1$, (\ref{2.3.3e}), and taking into account the
relations between coefficients $C_p$ and $D_p$ both for the $HE_1$ and $HM_1$ waves, for the $p$-th 
harmonic in the total deflecting field we get:
\begin{equation}
e_{dxp}(x,y) \approx \frac{i k_{sp}}{2}(1+\frac{(k^*_{sp} x)^2+(k^*_{sp} y)^2}{4})(A+B)e_{zp} ,\quad e_{dyp}(x,y) \approx 0.
\label{3.2.6e} 
\end{equation} 
Here the spatial harmonic $e_{zp}$ for the total $E_z$ component is used. In the derivation 
of (\ref{3.2.5e}) the transformations are the same for the $HE_1$ and the $HM_1$ component and $C_p-D_p = 
\frac{k_{zp}+k}{ k_{sp}}$ for both waves.\\  
The relation (\ref{3.2.6e}) indicates the possibility that for $A \approx -B$ the amplitude of the $p$-th 
spatial harmonic $e_{dxp}(x,y)$ in the deflecting field $E_d$ reduces regardless of the amplitude of the $p$-th harmonic  
in the distributions of the original field component $e_{zp}$. \\
As one can see, comparing the expressions for the synchronous harmonic $p=0$ for $\beta <1$ (\ref{3.1.4e}) and 
for $p \neq 0$, (\ref{3.2.6e}), it looks very similar. But there is a big difference. The higher
spatial harmonics do not vanish, even for $\beta=1$. It have a field modulation in the $z$ direction, reflected in 
(\ref{3.2.6e}) by the first multiplier $k_{sp}$,  for a synchronous particle. The 
constant (with respect $x,y$) term in the transverse direction in $e_{dxp}$ is modulated in the $z$ direction. 
As one can estimate from (\ref{2.2e}), (7), for $|p| > 1, |k_{sp}| \sim \frac{2 \pi p}{d} \gg k$. 
Non linear additions in $e_{dxp}$ rise fast with $r$, as $\sim (\frac{2 \pi p r}{d})^2$.\\     
The higher spatial harmonics are the main source of non linear additions in the deflecting field
distribution. 
\subsection{Multipole additions}
A complete rotational symmetry is not allowed for operating a DS. In this case two waves with deflection 
in $x$ and in $y$ directions are degenerated in frequency, resulting in not predictable direction of the 
actual deflection. To cancel the degeneration, in a DLW special holes in the disks are used, which deteriorate the
axial symmetry. Also there are DS's, see for example Fig. 1d, which originally have no axial symmetry.\\
But to support the deflecting mode with $n=1$, the DS geometry should be symmetric.
The deflection in $x$ direction corresponds to zero normal magnetic field at the plane $x=0, H_{norm}=0$. 
And the deflecting mode satisfies the boundary condition of zero tangential electric field $E_{\tau}=0$ at the 
plane $y=0, E_{\tau}=0$.  But these boundary conditions simultaneously satisfy also waves with $n=3,5,7,9 ...$.\\
In the field distribution of a real DS, especially with essential deterioration of the rotational symmetry, 
similar to the DS shown in Fig. 1a, always such higher multipole components $n =3,5,7...$ with a
dependence on the azimuth as (\ref{2.1.3e}) are present. Similar to spatial harmonics in the deflecting mode, the multipole 
field components in the beam aperture are required to satisfy the boundary conditions at $r=a$ and decay 
towards the beam axis. These components should be presented in (\ref{2.2e}) with 
additional summations over $n$. But we can consider the properties of these components independently,
 due to the linearity of the Maxwell equations.\\
Let us consider the deflecting field from the first multipole component - the synchronous sextupole 
wave $n=3$ for $\beta=1$. 
Taking the distribution of the $E_z$ component from Table 2 and using (\ref{3.6e}) we find: 
\begin{equation}
E_z = A_0 \frac{k r^3}{48} cos(3 \vartheta), \quad \vec F_d = \frac{e A_0}{16} r^2 (\vec i_r cos(3 \vartheta)-  \vec i_{\vartheta} sin(3 \vartheta)) 
\label{3.3.1e}
\end{equation}
Transferring to Cartesian coordinates, 
\begin{eqnarray}
\vec F_d = \frac{e A_0}{16} (\vec i_x (x^2-y^2) + \vec i_y 2 xy).
\label{3.3.2e}
\end{eqnarray}
The transverse force from multipole components even for the synchronous harmonics has only non linear 
terms. They are non vanishing synchronous additions. The amplitudes of multipole additions depend on 
the DS geometry and should be minimized, as far as possible, during the DS shape development.\\
Similar to the dipole mode, we can consider the spatial harmonics for multipole waves. Always it will 
be just nonlinear additions, starting with higher power than $x^2, y^2$. The higher spatial harmonics
in the multipole component decay faster toward the beam axis and are hence not so dangerous for the field quality as 
synchronous multipole waves. 
\subsection{Bunch deflection and bunch rotation}
Let us consider the difference between bunch deflection and bunch rotation for a TW  
operating mode.\\
The force from the field component $E_{j}(\vartheta,r,z)$ in (1) on the particle, moving with 
velocity $\beta c$, is: 
\begin{equation}
\frac{F_{j}(\vartheta,r,z)}{e}= \Re \widehat{E_j(\vartheta,r,z)} e^{i( \psi_j (\vartheta,r,z)+ k_{z0} z+\phi)}, \quad 
\omega t = \frac{\omega z}{\beta c}= k_{z0}z, 
\label{4.1e}
\end{equation}
where $\phi$ is the initial phase shift between the particle and the wave.\\
For bunch deflection $\phi=0$ for the central particle and, using (\ref{3.4e}) for the
deflecting field, from (\ref{4.1e}) it follows:
\begin{equation}
\frac{F_{dr,d\vartheta}(\vartheta,r,z)}{e}=  a_{d0,dr,d\vartheta}(\vartheta,r) +
 \sum_{p =1} a_{dp,dr,d\vartheta}(\vartheta,r) cos(\frac{ 2 p \pi z}{d}). 
\label{4.2e}
\end{equation}
The particles get a permanent deflection from the synchronous interaction with the main harmonic $p=0$ 
and the bunch oscillates as a whole with respect of the deflection direction. Usually the main spatial harmonic 
dominates in the field expansion and the effect of higher spatial harmonics can be acceptable for bunch 
deflection – the bunch oscillation is in the background of the dominating deflection.
 In bunch deflection mode exists an analogy with the particle acceleration due to  
a longitudinal field - there is a permanent acceleration from the synchronous harmonic and oscillations with respect to the
synchronous motion due to the interaction with higher spatial harmonics.\\
For the general description of an accelerating field in accelerating structures the transit time factor $T_z$ 
is used and we can define in an equivalent way the transit time factor $T_d$ for the deflecting field at the DS axis:
\begin{equation}
T_z = \frac{|\int_{-\frac{d}{2}}^{\frac{d}{2}} E_z(z) e^{ik_{z0}z} dz|}
{\int_{-\frac{d}{2}}^{\frac{d}{2}} \widehat{E_z(z)} dz},\quad
T_d = \frac{|\int_{-\frac{d}{2}}^{\frac{d}{2}} E_d(z) e^{ik_{z0}z} dz|}
{\int_{-\frac{d}{2}}^{\frac{d}{2}} \widehat{E_d(z)} dz}
\label{4.3e}
\end{equation} 
\begin{figure}[htb]
\centering
\epsfig{file=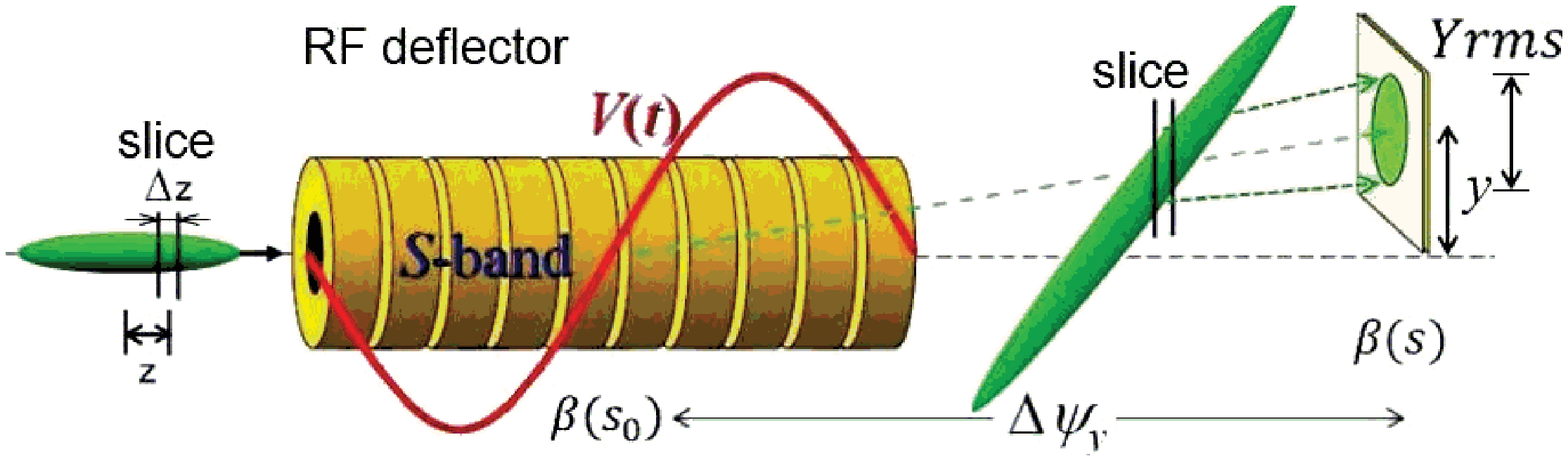, width =130.0mm}
\begin{center}
 Figure 2: Illustration of the bunch rotation for measurements of longitudinal distributions, \cite{malutin}.
\end{center}
\label{2f}
\end{figure} 
For bunch rotation $\phi=\frac{\pi}{2}$ and, similar to (\ref{4.2e}),
\begin{equation}
\frac{F_{dr,d\vartheta}(\vartheta,r,z)}{e}=  \sum_{p =1} a_{dp,dr,d\vartheta}(\vartheta,r) sin(\frac{ 2 p \pi z}{d}). 
\label{4.4e}
\end{equation}
Schematically a bunch rotation is illustrated in Fig. 2, see \cite{malutin} for more explanations about measurements of bunch 
parameters. During bunch rotation the central bunch particle doesn't see the main spatial harmonic
and moves along the DS axis. But upstream and downstream particles will receive a synchronous deflection in opposite 
directions - the bunch rotates. Together with the synchronous rotation, the bunch oscillates as a whole with respect to the
DS axis due to the higher spatial harmonics in (\ref{4.4e}).\\
In structures with planes of mirror symmetry the effective deflecting field in (\ref{4.2e}) is an even 
function on $z$ with respect to the mirror planes and the residual field in (\ref{4.4e}) is an odd function.
\subsection{Criterion for higher harmonics estimation in periodical structures}
 To compare different DS realizations with respect to
the relative level of higher harmonics in the field distributions, we need a criterion for this comparison.\\
Let us consider the phase dependence $\psi_j (\vartheta,r,z)$ in (1). Subtracting the synchronous spatial harmonics from the phase distribution, from (\ref{2.2e}) we get:
\begin{equation}
E_{j}(\vartheta,r,z)=\widehat{E_j(\vartheta,r,z)} e^{i \psi_j (\vartheta,r,z)+k_{z0} z}=
\sum^{p \rightarrow + \infty}_{p \rightarrow - \infty} a_{jp}(\vartheta,r) e^{-i \frac{2p \pi z}{d}},
\label{4.1.1e}
\end{equation}
or 
\begin{eqnarray}
\Re \widehat{E_j(\vartheta,r,z)} e^{i \psi_j (\vartheta,r,z)+k_{z0} z}=
\sum _{p=0} a_{jp}(\vartheta,r) cos( \frac{2p \pi z}{d}),\\
\nonumber
\Im \widehat{E_j(\vartheta,r,z)} e^{i \psi_j (\vartheta,r,z)+k_{z0} z}=
- \sum_{p =1} a_{jp}(\vartheta,r) sin( \frac{2p \pi z}{d}).
\label{4.1.2e}
\end{eqnarray}
In (42) we find a simple expansion in a Fourier set, where 
\begin{equation}
a_{jp}(\vartheta,r) = \frac{1}{d} \int_{-\frac{d}{2}}^{\frac{d}{2}} E_{j}(\vartheta,r,z) sin( \psi_j (\vartheta,r,z)+k_{z0} z) sin(\frac{2p \pi z}{d})dz.
\label{4.1.3e}
\end{equation}
Spatial harmonics are essential at the aperture radius $r=a$ and we can assume 
$a_{jp}(\vartheta,a) \sim a_{0p}(\vartheta,a)$. As one can see from (14) and (15), the 
distributions for harmonics are described with combinations of Bessel functions. According to (\ref{2.2.2e}), 
the higher harmonics $|p| \gg 1$  attenuate to the axis as: 
\begin{equation}
a_{jp}(0) \sim a_{jp}(a) \cdot exp(-\frac{4 \pi ^2 |p|}{ \beta \Theta_0} \cdot \frac{a}{\lambda} ). 
\label{4.1.4e}
\end{equation} 
At the beam axis $r=0$ just the lower harmonics $p=\pm 1, \pm 2, \pm 3 $ are really essential.\\
For an 'in total' estimation of the harmonic content we will consider the deviation of the total phase distribution 
$\psi_j (\vartheta,r,z)$ at the axis from the phase of the synchronous harmonic $\delta \psi_j (z)$ and use the parameter 
 $\Psi_j$ 
\begin{equation}
\delta \psi_j(z) =  \psi_j(z) +\frac{\Theta_0 z}{d},\quad
\Psi_j  = max(|\delta \psi_j(z)|),\quad 0 \leq z \leq d, r=0. 
\label{4.1.5e}
\end{equation} 
From symmetry properties $\delta \psi_j(z)$ is always an odd function with respect to the mirror symmetry planes and
we can use only half of period length in (\ref{4.1.5e}). For the cases of our interest $|a_{jp}| \ll |a_{0p}|$ 
it is $\sim sin(\frac{2 pi z}{d})$ function with small deviations due to the second and the third harmonics. 
In this case the usage of $\Psi_j$ instead of $\delta \psi_j(z)$ in (\ref{4.1.3e}) will lead to an upper estimation of $a_{jp}(0)$
and smaller values of $\Psi_j$ correspond to smaller values of $a_{jp}(0)$.\\
The slow wave system with perfectly linear phase dependence $\psi_j (\vartheta,r,z)$ is not possible, because 
it means $\delta \psi_j(z)=0$ and, from (\ref{4.1.3e}),  $a_{jp}=0, |p| \neq 0$ corresponds to a single 
wave, which can exists only in a smooth waveguide. But slow wave systems with essentially damped 
harmonics at the beam axis are possible.
\subsection{Panofsky-Wenzel theorem} 
The relationship
\begin{equation}
\vec p_t = -i (\frac{ec}{\omega}) \int_0^L \nabla_t E_z dz
\label{5.1e}
\end{equation}
where $\vec p_t$ is the transverse particle momentum gain, has been derived in \cite{pif} regardless of 
the field classification, but with an important assumption – the particle velocity is large enough 
to allow the particle direction to remain essentially unchanged by the transverse force. It is the 
case $\beta=1$. For lower electron energy this statement is a framework.\\
As one can see comparing (\ref{3.6e}) and (\ref{5.1e}), both formulations lead to similar value of the deflecting 
force. But in (\ref{5.1e}) the total value of the $E_z$ component, without specification of the 
synchronous space harmonic, is used.\\
\begin{figure}[htb]
\centering
\epsfig{file=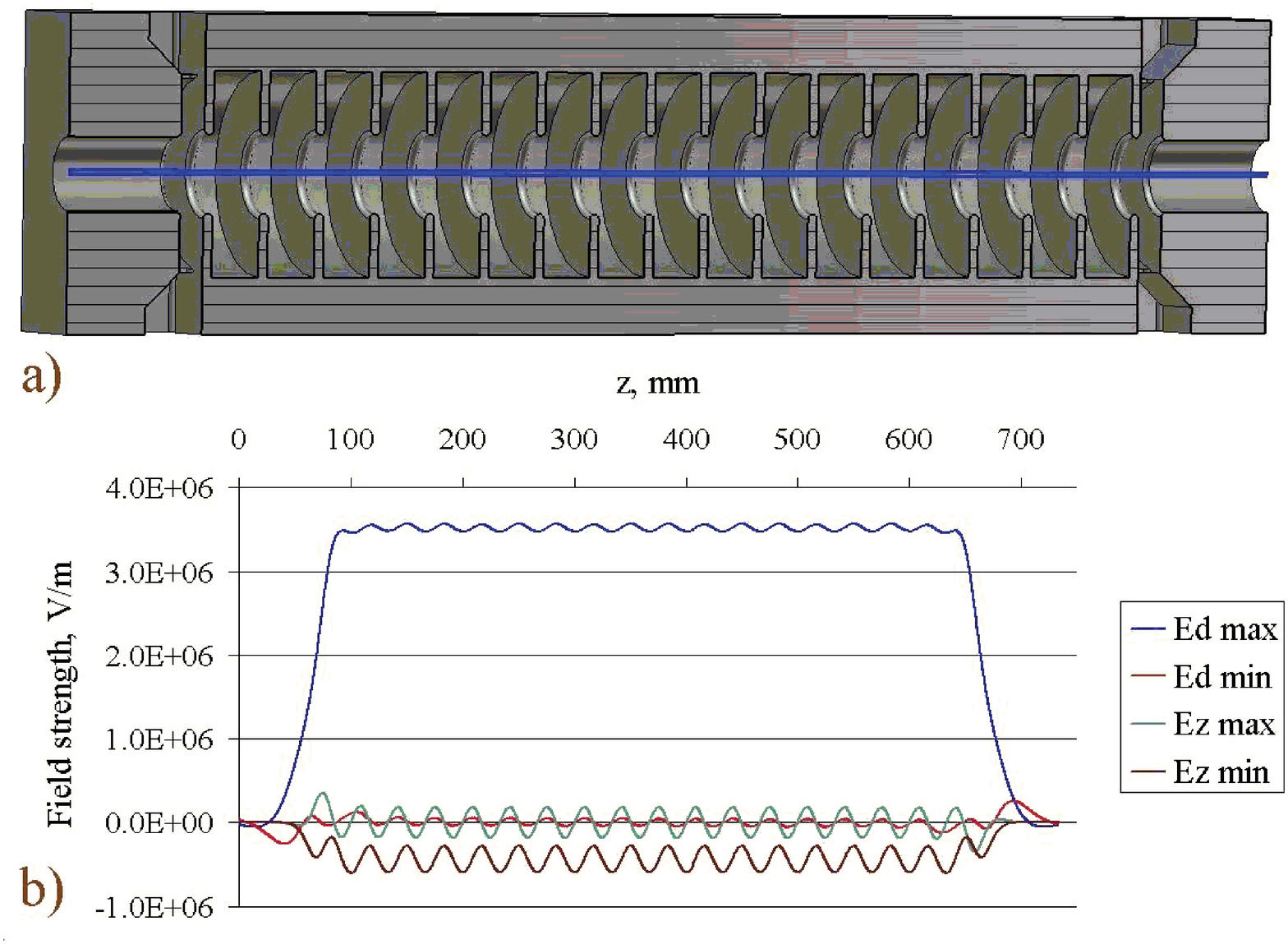, width =140.0mm}
\begin{center}
 Figure 3: A TW DLW structure (a) with $\Theta_0 =\frac{2\pi}{3}$ and field distributions 
for $E_d$ at the axis, $\phi_0 = 0$ (dark blue) and $\phi_0=\frac{\pi}{2}$ (red), and $E_z$ distribution along the line  
$x=2 mm$ also for $\phi_0 = 0$ (blue) and $\phi_0=\frac{\pi}{2}$ (brown).
\end{center}
\label{3f}
\end{figure}    
From the Maxwell equation $rot \vec E = -\frac{\partial \vec B}{\partial t}$ we find directly:
\begin{equation}
\frac{\partial E_z}{\partial r} = \frac{\partial E_r}{\partial z} + i k Z_0 H_{\vartheta} \quad or \quad 
\frac{\partial E_z}{\partial x} = \frac{\partial E_x}{\partial z} + i k Z_0 H_y. 
\label{5.2e}
\end{equation}
We can expand each component either into a Fourier series over space harmonics, (\ref{2.2e}) (discrete $k_{zp}$ 
spectrum for periodical structures) or provide a Fourier transform - continuous $k_{zp}$ 
spectrum for single cavity. The interaction with a particle, traveling along $z$ with unchanged 
velocity, is described by $e^{i k_{z0} z}$ and will select only the synchronous component. With the inverse transformation, we obtain:
\begin{equation}
 \frac{\partial e_{z0}(r)}{\partial r} = -i k_{z0}( e_{r0}(r) + \beta_0 Z_0 
h_{\vartheta 0}(r)) =-ik_{z0} e_{d0}(r),
\label{5.3e}
\end{equation} 
where $e_{d0}$ is the amplitude of the synchronous harmonic of the equivalent deflecting field, see (\ref{3.2e}).\\  
As one can see from expressions for $HM-HE$ wave components in the Table 2, for a dipole mode  
$E_z(r) \sim kr, \Rightarrow \frac{\partial E_z(r)}{\partial r} \sim k$.
Instead of the representation via the total $E_z$ component, the coupling of longitudinal and transverse motion 
is generated through the synchronous harmonics.\\ 
This theorem provides an important indication – the longitudinal and transverse forces are shifted in phase 
at $\frac{\pi}{2}$.\\
 In Fig. 3a a TW deflecting structure with operating phase advance $\Theta_0=\frac{2\pi}{3}$ is shown. 
This structure is the well known LOLA IV option, \cite{lola}, scaled to an operating frequency of $3000 MHz$ 
and differing in the design of the RF couplers. In Fig. 3b the distributions of $E_d$ and 
$E_z$ are shown for an operation in bunch deflection ($\phi_0=0$) and bunch rotation mode ($\phi_0=\frac{\pi}{2}$). 
The plots in Fig. 3b illustrate - for bunch deflection $E_d$ is maximal, but the average 
$E_z$ is zero and there are only oscillations due to $E_z(z)$, similar to (\ref{4.4e}). The bunch deflection 
is not accompanied by an average change in the longitudinal momentum. For bunch rotation $E_d$
is zero on average, according (\ref{4.4e}), while $E_z$ is on average not zero, 
similar to (\ref{4.2e}). The bunch rotation is accompanied by an average change in the longitudinal 
particle momentum which is proportional to radial particle position. 
\section{Traveling and standing wave operation}
Let us consider some differences due to different operating modes of DS.
\subsection{Deflecting field distributions for TW and SW mode}
\begin{figure}[htb]
\centering
\epsfig{file=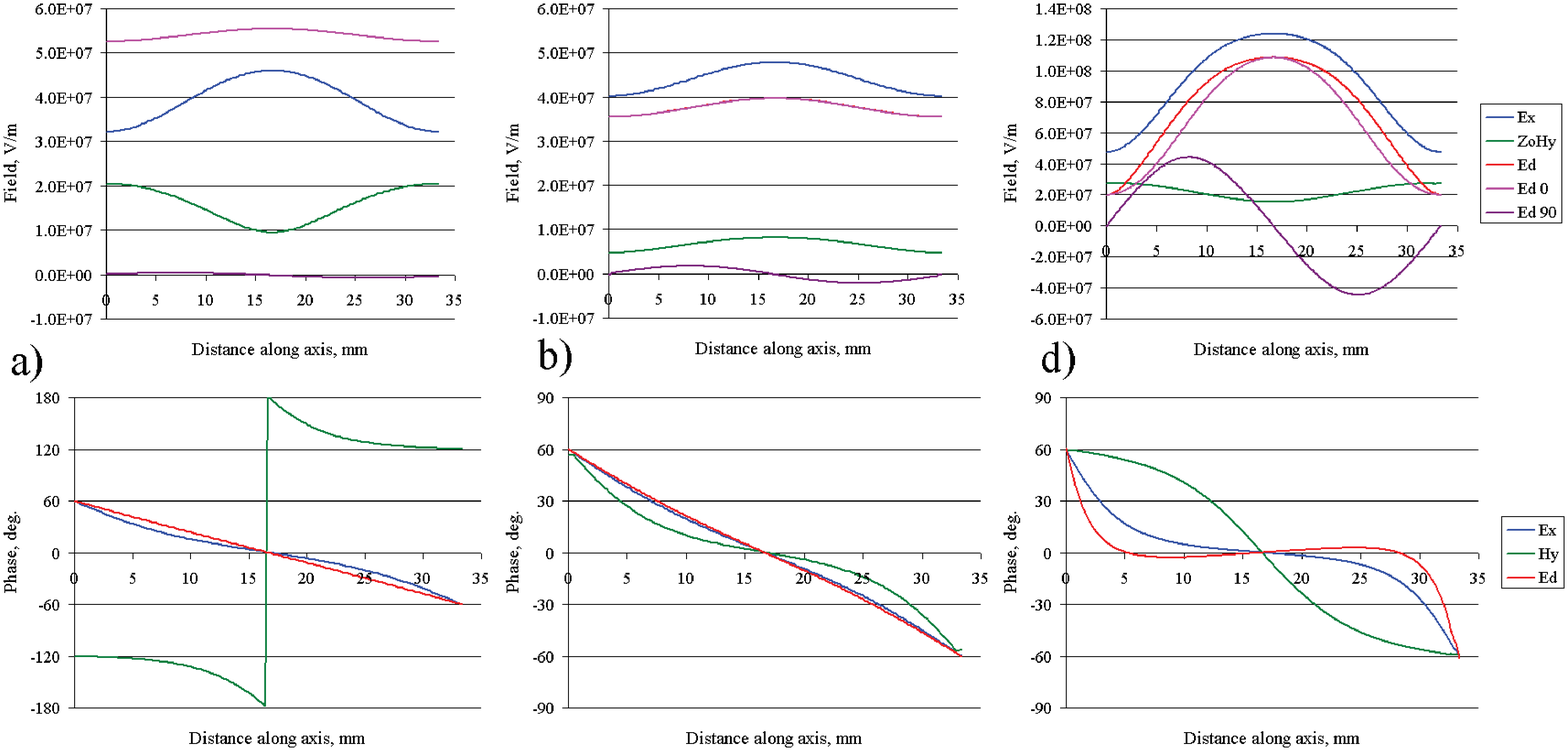, width =150.0mm}
\begin{center}
 Figure 4: The distributions of amplitudes for $E_x$, $Z_0 H_y$ and deflecting field 
$E_d, \phi=0$ and $E_d, \phi=90$ (upper row) and distribution of phases for $E_x$, $H_y$ and $E_d$ 
(bottom row) for DS's, shown in Fig. 1a, b and d, TW operating mode, $\Theta_0 =\frac{2 \pi}{3}$.
\end{center}
\label{4f}
\end{figure}
For a TW mode each field component has both real and imaginary parts. The amplitudes of all 
spatial harmonics in (\ref{2.2e}) are constant over the structure period $d$. 
Considering the deflecting field composed as (\ref{3.2e}) or (\ref{3.1.5e}), for the deflecting force,
similar to (\ref{4.1e}), we have at the DS axis:
\begin{eqnarray}
\frac{F_d(z)}{e} = \Re \bigl( \sum_{p=0}(e_{rp}-Z_0 h_{\vartheta p}) e^{i( k_{zp}+k_{z0} z+\phi)} \bigr )
= (e_{r0}-Z_0 h_{\vartheta 0}) cos(\phi) +\\
\nonumber
 +cos(\phi) \sum_{p =1} (e_{rp}-Z_0 h_{\vartheta p})
cos(\frac{2 p \pi z}{d})+sin(\phi) \sum_{p =1} (e_{rp}-Z_0 h_{\vartheta p})sin(\frac{2 p \pi z}{d}).            
\label{6.1.1e}
\end{eqnarray}
In Fig. 4 the distribution of the amplitudes for $E_r=E_x$, $Z_0 H_{\vartheta} =  Z_0 H_y$ 
and the  deflecting fields 
$E_d, \phi=0$ and $E_d, \phi=\frac{\pi}{2}$ (upper row) and the distribution of the phase for $E_x$, $H_y$ and $E_d$ 
(bottom row) for the DS, shown in Fig. 1a, b and d, assuming a TW operating mode with 
$\Theta_0 =\frac{2 \pi}{3}$, are shown. There are some differences in deflecting field distributions due 
to different phasing and balance of hybrid waves $HE_1$ and $HM_1$ in (\ref{2.3.3e}), but the common 
properties are the same:\\
The interaction with the synchronous harmonic is constant over the period, regardless to the ratio 
of $e_{r0}$ and $Z_0 h_{\vartheta 0}$ components. Always an initial phase shift $\phi=\frac{\pi}{2}$ between the wave 
and the central particle of the bunch exists when the central particle  doesn't see the synchronous harmonic – i.e. 
 the perfect bunch rotation operating mode. Always oscillations of $E_d$ take place due to higher spatial harmonics only.\\ 
For TW mode harmonics the attenuation (\ref{4.1.4e}) works for all original field components and 
is especially essential for lower phase advance $\Theta_0 \ll \pi$. Every DS, operating in TW mode 
with a low phase advance, has a reduced level of spatial harmonics, both for the longitudinal 
and transverse field components. For the deflecting field $E_d$, composed from electric and magnetic 
components of the original field, further reduction of harmonics takes place for opposite phasing of $HE_1$ 
and $HM_1$ waves, $A \cdot B < 0$ in (\ref{2.3.3e}). According to (\ref{3.2.6e}), we can expect $|e_{dp}|
 \ll |e_{zp}|$ for $A \sim -B$.\\
The criterion for harmonics estimation (\ref{4.1.5e}) in TW mode works naturally both for 
longitudinal and transverse components of the Lorenz force (\ref{3.1e}). The minimal value of $\Psi_d$ 
corresponds to a minimal level of spatial harmonics for bunch rotation $\phi=\frac{\pi}{2}$ and, 
simultaneously, for bunch deflection $\phi=0$, according to (\ref{4.1.2e}).\\
For SW operation mode the field distribution for $ 0 < \Theta_0 < \pi$ can be obtained as the sum of the 
forward wave $E_{j}(\vartheta,r,z)$, (\ref{2.1e}), and the backward wave $E_{j}^*(\vartheta,r,z)$. But for 
$\Theta_0 =0, \pi$ the forward and the backward waves are identical, \cite{slow}. A specific case of a 
compensated structures is, see \cite{lapost}, when at the operating frequency two $0$ or two $\pi$ modes from 
two different passbands, which have a conjugated parity of field distributions with respect to mirror symmetry planes, coincide. 
There is just one proposal of the compensated DS known, \cite{comp_defl}, which is not realized in practice.\\
As for every periodical structure, SW modes for DS with $ 0 < \Theta_0 < \pi$ are not effective with respect to the RF parameters. 
There are no references for DS, operating with $\Theta_0=0$ or $\Theta_0=2 \pi$. Considering DS SW 
operation, we assume in the following $ \Theta_0 = \pi$.\\
\begin{figure}[htb]
\centering
\epsfig{file=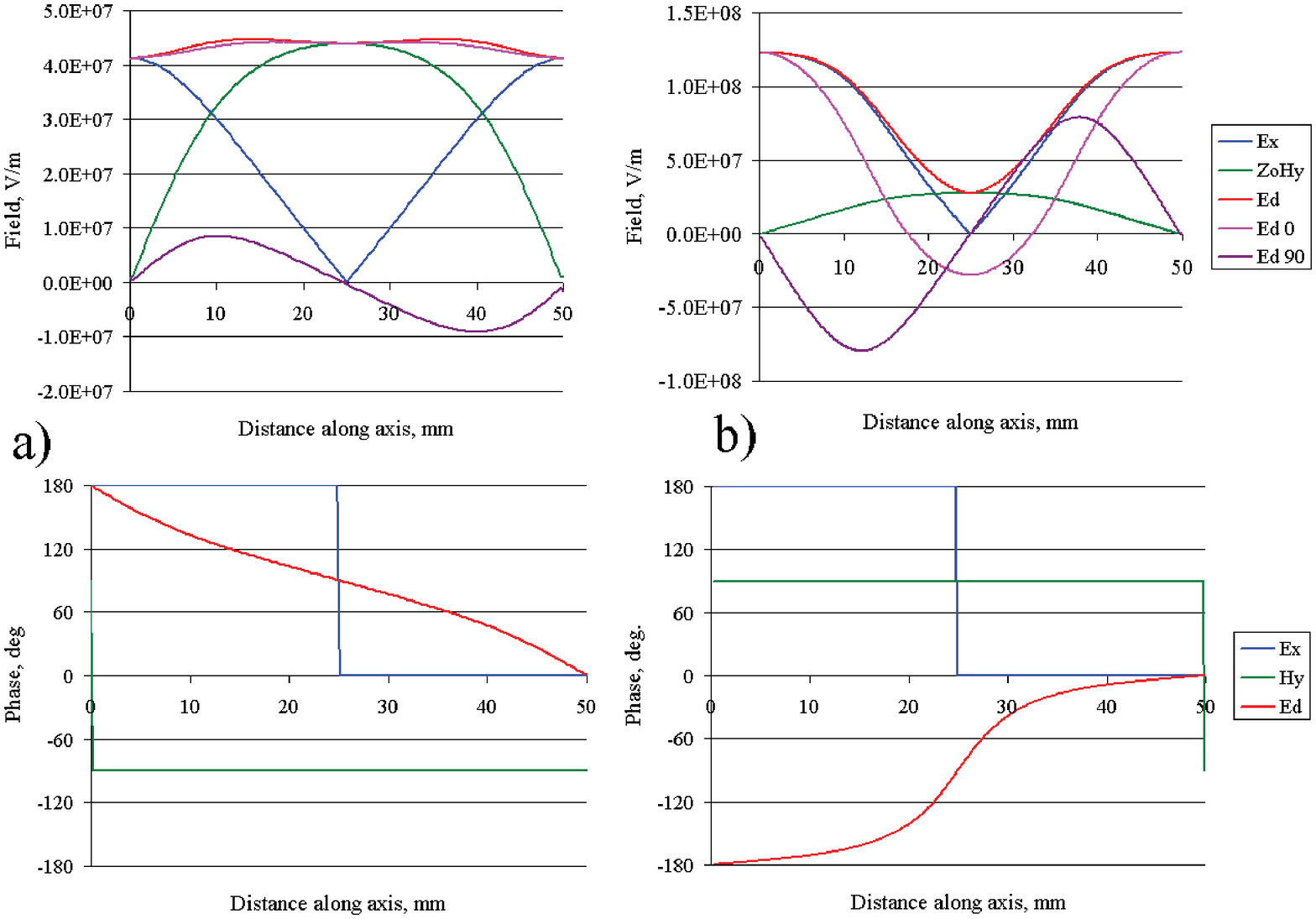, width =110.0mm}
\begin{center}
 Figure 5: The distributions of amplitudes for $E_x$, $Z_0 H_y$ and $E_d$, deflecting field 
$E_d, \phi=0$ and $E_d, \phi=90$ (upper row) and distribution of phase for $E_x$, $H_y$ and $E_d$ 
(bottom row) for DLW with $\frac{a}{\lambda} =0.18$, (a) and for decoupled DS, Fig. 1d, 
$\frac{a}{\lambda} =0.07,\quad \Theta_0 = \pi$.
\end{center}
\label{5f}
\end{figure} 
For the SW case each field component $E_{j}(\vartheta,r,z)$ in (1) has only either real or 
imaginary part and the phase $\psi_j (\vartheta,r,z)$ is a step-wise function with one step 
of $\pi$ over the period. The periodicity in the field distribution, similar to (\ref{2.2e}) and the 
linearity in (\ref{3.2e}) remain.\\ 
For each component separately $\Psi_j =\frac{\pi}{2}$ holds, according to (\ref{4.1.5e}).
The longitudinal component in the Lorenz force (\ref{3.1e}) is only due to the $E_z$ component and a reduction of harmonics 
can be obtained only due to attenuation (\ref{4.1.4e}), which is not so effective for lower harmonics 
at $\Theta_0=\pi$. Typical values of $T_z$, (\ref{4.3e}), are $T_z \approx (0.75 \div 0.8), 
e_{z 1} \sim 0.6$. A reduction of aberrations in the longitudinal force component for the SW mode 
can be achieved only at the expense of an increased aperture radius.\\ 
In Fig. 5 the distributions of amplitudes for $E_r=E_x$, $Z_0 H_{\vartheta} =  Z_0 H_y$ 
and deflecting fields $E_d, \phi=0$ and $E_d, \phi=\frac{\pi}{2}$ (upper row) and distributions of phase for $E_x$, $H_y$ 
and $E_d$ (bottom row) for DLW $\frac{a}{\lambda} =0.18$, (a) and for decoupled DS, Fig. 1d, 
$\frac{a}{\lambda} =0.07$ are shown for SW operation mode.\\ 
Taking into account DS symmetry properties, for deflecting field $E_{dr}$ we can write:
\begin{equation}
E_{dr}(z)= cos(\omega t + \phi) \cdot  \sum_{p = 0} e_{rp} cos(\frac{(\pi+ 2 p \pi) z}{d}) -
sin(\omega t + \phi) \cdot \sum_{p = 0} Z_0 h_{\vartheta p} sin(\frac{(\pi+ 2 p \pi) z}{d}).
\label{6.1.2e}
\end{equation}
Considering just the main harmonics $p=0$ for the synchronous particle $\omega t = k z = \frac{\pi z}{d}$, we see:
\begin{eqnarray}
E_{dr}(z) = e_{r0} cos(kz + \phi) cos (kz)  - Z_0 h_{\vartheta 0} sin(kz + \phi) sin (kz) =\\
\nonumber
= \frac{e_{r0} -  Z_0 h_{\vartheta 0}}{2} \cdot cos(\phi) + \frac{e_{r0} +  Z_0 h_{\vartheta 0}}{2} \cdot cos( 2 kz+ \phi)\bigr )
\label{6.1.3e}
\end{eqnarray}
For any intial phase shift $\phi$ the central particle sees both a uniform and an oscillating impact of the deflecting 
field. Even if the main harmonics are free from aberrations, the oscillating part in (51) 
shifts the particle from the DS axis to regions with higher field nonlinearities from the spatial harmonics. 
As one can see from (51), the ratio of uniform and oscillating parts depends on the 
phasing and the balance of the synchronous harmonics $e_{r0}$ and $Z_0 h_{\vartheta 0}$. For opposite 
phasing $e_{r0} \cdot Z_0 h_{\vartheta 0} <0 $ the amplitude of the uniform deflection $e_{r0} -  Z_0 h_{\vartheta 0}$ exceeds 
the amplitude of the oscillations $e_{r0} +  Z_0 h_{\vartheta 0}$. For the case $e_{r0} = -  Z_0 h_{\vartheta 0}$, 
corresponding to $A=-B$ in (\ref{2.3.3e}), the central particle moves under action of the synchronous 
field harmonics without oscillations and the perfect bunch rotation is possible for $\phi=\frac{\pi}{2}$.\\  
Taking into account the spatial harmonics in the $E_r$ and $H_{\vartheta}$ field components, for the deflecting 
field one can get:
\begin{eqnarray}
E_{dr}(z) = \frac{e_{r0} -  Z_0 h_{\vartheta 0}}{2} \cdot cos(\phi) +\\
\nonumber
+ \frac{cos \phi }{2} \cdot \sum_{p = 1}(e_{r p-1} +  Z_0 h_{\vartheta p-1}+ e_{rp} -  Z_0 h_{\vartheta p})cos (\frac{2p \pi z}{d}) - \\
\nonumber 
- \frac{ sin \phi }{2} \cdot \sum_{p = 1}(e_{r p-1} + Z_0 h_{\vartheta p-1} - e_{rp} +  Z_0 h_{\vartheta p})sin (\frac{2p \pi z}{d}) \big ).
\label{6.1.4e}
\end{eqnarray}
This is the general case of the force onto the central particle from the force sources, $E_r$ and $H_{\vartheta}$, 
which are shifted both in distance along the axis and in time. As one can see the amplitudes 
of the $p$-th and $p-1$ harmonics are coupled in the amplitudes
of the oscillations $ \sim  sin (\frac{2p \pi z}{d}), \sim cos (\frac{2p \pi z}{d})$. Just for the case $e_{r0} = -  Z_0 h_{\vartheta 0}$ 
oscillations appear due to the higher spatial harmonics only. For equal phasing $e_{r0} \cdot Z_0 h_{\vartheta 0} > 0 $ 
the amplitude of the oscillations can exceed the average value of deflecting field.\\
We can rewrite (52) as:
\begin{eqnarray}
E_{dr}(z) = \frac{e_{d0}^{'}}{2}  \cdot cos\phi 
+ \frac{ cos \phi }{2} \cdot \sum_{p = 1}e_{d p}^{e} cos (\frac{2p \pi z}{d}) 
- \frac{sin \phi }{2} \cdot \sum_{p = 1}e_{d p}^{o} sin (\frac{2p \pi z}{d}).\\
\nonumber
e_{d0}^{'}=e_{r0} -  Z_0 h_{\vartheta 0},\quad
 e_{d p}^{e}=e_{r p-1} +  Z_0 h_{\vartheta p-1}+ e_{rp} -  Z_0 h_{\vartheta p},\\
\nonumber
 e_{d p}^{o}=e_{r p-1} + Z_0 h_{\vartheta p-1} - e_{rp} +  Z_0 h_{\vartheta p}.
\label{6.1.4e1}
\end{eqnarray}
As one can see from (53), the amplitudes of the even oscillating terms $ e_{d p}^{e}$ differ 
from the amplitudes of the odd oscillating terms $ e_{d p}^{o}$ for the same $p$.\\
The criterion for harmonics estimation (\ref{4.1.5e}) in SW mode works formally as for a TW mode.
Taking into account the time dependence, for deflecting field we can write: 
\begin{equation}
E_{dr}(z)e ^{i \omega t} = (E_r (z) - i Z_0 H_{\vartheta} (z)) e ^{i \omega t}\quad or \quad 
E_{dr}(z)e ^{i kz} = (E_r (z) - i Z_0 H_{\vartheta} (z)) e ^{i kz}  .
\label{6.1.5e}
\end{equation}
and can consider $E_{dr}(z)e ^{i kz}$ as a complex function with real and the imaginary parts, 
which describes an  'equivalent' traveling wave. In a mathematical sense the parameter $\Psi_d  = max(|\delta \psi_d(z)|)$ 
reflects the imaginary part of the complex function (\ref{6.1.5e}). For the opposite $e_{r0}, h_{\vartheta 0} $ phasing 
$e_{r0} \cdot Z_0 h_{\vartheta 0} < 0 $ the parameter $\Psi_d$ reflects 'in total' the oscillating part in 
(\ref{6.1.4e1}), which is described by the odd oscillating terms $ e_{d p}^{o}$, including both higher spatial 
harmonics and oscillations due to the not compensated term  $e_{r0} +  Z_0 h_{\vartheta 0} \neq 0$ in the main 
harmonics. Thus the condition $(\Psi_{d} )_{min} $ corresponds to the most 
uniform motion of the central particle near the axis during bunch rotation for a SW mode. But the $E_d$ distribution 
for the bunch deflection condition $(\Psi_{d} )_{min} $ corresponds due to the $ e_{d p}^{o}$ and $ e_{d p}^{e}$ difference not to the minimal level of oscillating 
additions. In the distribution $E_d(z), \phi=0$, 
obtained for the condition $(\Psi_{d} )_{min} $ we will always have larger deviations from the average value, than 
in the distribution $E_d(z), \phi=\frac{\pi}{2}$.\\
For equal $e_{r0}, h_{\vartheta 0} $ phasing  the parameter $\Psi_d$ is not useful, 
because the 'equivalent' traveling wave in (\ref{6.1.5e}) is the backward wave with respect to the central 
particle and $(\Psi_d )_{min}  = \pi$. But for equal phasing 
we have both strong oscillations due to the main harmonics, according to (\ref{6.1.3e}), and increased 
amplitudes for higher $E_d$ harmonics, according to (\ref{3.2.6e}).    
\subsection{Phase deviations in deflecting field distribution}  
Suppose for a TW mode we have at the DS axis a phase deviation $d \psi_d (z)$ in the phase distribution $\psi_d (\vartheta,r,z)$ 
of the deflecting field due to some reasons.  
Similar to (\ref{4.1e}) and (\ref{4.2e}), the force from the deflecting field for bunch deflection is: 
\begin{eqnarray}
\frac{F_d(z)}{e} = \Re \bigl(\widehat{E_d(z)} e^{i( \psi(z)+k_{z0} z+d \psi_d (z))} \bigr )=  \widehat{E_d(z)} cos(\delta \psi_d(z)+d\psi_d (z)) =\\
\nonumber 
= cos( d \psi_d (z)) \cdot ( e_{d0} + \sum_{p =1} e_{dp} cos(\frac{ 2 p \pi z}{d})) -sin( d \psi_d (z)) \cdot  \sum_{p =1} e_{dp} sin(\frac{ 2 p \pi z}{d}).
\label{6.2.1e}
\end{eqnarray}
where (\ref{4.2e}) and (\ref{4.4e}) are taken into account. 
For bunch rotation 
\begin{eqnarray}
\frac{F_d(z)}{e} = \Re \bigl(\widehat{E_j(z)} e^{i( \psi(z)+k_{z0} z+d \psi_d (z)+\pi /2)} \bigr )= 
 \widehat{E_d(z)} sin(\delta \psi_d(z)+d\psi_d (z)) =\\
\nonumber 
= sin( d \psi_d (z)) \cdot ( e_{d0} + \sum_{p=1} e_{dp} cos(\frac{ 2 p \pi z}{d})) +cos( d \psi_d (z)) \cdot  \sum_{p =1} e_{dp} sin(\frac{ 2 p \pi z}{d}).
\label{6.2.2e}
\end{eqnarray}
From (55) and (56) we see a quite different role of the phase deviation $d \psi_d (z)$ 
for bunch deflection and bunch rotation. For small deviations $|d \psi_d (z)| \ll 1$
\begin{equation}
cos (d \psi_d (z)) \approx 1 - \frac{(d \psi_d (z))^2}{2}, \quad sin(d \psi_d (z)) \approx d \psi_d (z).
\label{6.2.3e}
\end{equation} 
For bunch deflection the phase deviations $d \psi_d (z)$ leads to a second order change in the 
distribution of the deflecting field (\ref{4.2e}) and to the generation of small residual field (\ref{4.4e}).
For a bunch rotation mode the phase deviations immediately generate a bunch deflection 
$\approx e_{d0} d \psi_d (z)$ by the synchronous harmonic. At the plots of $E_d(z), \phi =\pi/2$ arise 
 peaks of $E_d$ and the bunch as a whole will be deflected from the axis, similar to the 
oscillations in (51), with the same effect. The bunch will be displaced as a whole from 
the axis to a region with higher level of nonlinear additions in the field distributions.\\ 
We have to distinguish possible reasons of phase deviations. 
The first reason are possible errors in the cell frequencies. It can be reduced by appropriate RF tuning.\\
The second reason are violations of the structure periodicity - RF coupler cell, end cell with connected 
beam pipe and so on. For such elements special attention is required.\\
The criterion for spatial harmonics estimation, introduced in (\ref{4.1.5e}), works well also 
for the total structure, including cells with violated periodicity. If we see large phase deviation 
in the RF coupler cell from the phase of the synchronous particle, according to (56) we will have 
corresponding unwanted peaks in the $E_d$ distribution for bunch rotation. Reduction of the phase deviation 
$d \psi_d (z)$ simultaneously leads to a reduction of the peaks.     
\subsection{RF efficiency for TW and SW mode}
Both for bunch deflection and bunch rotation we need the value of deflecting voltage $V_d$ for a  specified RF power $P$.\\
For SW mode all RF power is dissipated in the cavity surface and $V_d$ is directly related to the effective transverse 
shunt impedance $Z_e^{(SW)}$ per unit length:
\begin{equation}
Z_e^{(SW)} =\frac{|\frac{1}{k} \int_0^L  \frac{\partial E_z}{\partial z} e^{ik_{z0} z} dz|^2}{P_s L} 
=\frac{(E_{d0} L)^2}{P_s L},\quad \beta=1,
\label{6.3.1e}
\end{equation}
where $L$ is the total structure length. This definition of $Z_e$ is based on the Panofsky-Wenzel theorem.\\  
For TW mode the wave propagates along the structure, both
providing the required field and attenuating in amplitude due to RF power dissipation in the surface of the structure. 
The rest of the RF power comes to the RF load at the end of the TW structure. The TW structure parameter, 
independent with respect to frequency scaling, is the normalized field strength:
\begin{equation}
\frac{E_{d0} \lambda}{\sqrt{P_t}} = \sqrt{\frac{2 \pi\lambda Z_e^{(TW)}}{|\beta_g| Q}},
\label{6.3.2e}
\end{equation}
where $Z_e^{(TW)}$ is defined similar to (\ref{6.3.1e}), $E_{d0}$ is the field value for the 
synchronous deflecting harmonic and $P_t$ is the RF power flux. The total deflecting voltage 
$V_d^{(TW)}$ is:
\begin{equation}
V_d^{(TW)}=\int_0^L E_{d0}^{(i)} e^{-\alpha z} dz= \frac{E_{d0}^{(i)}}{\alpha} (1-e^{-\alpha L}),
\label{6.3.3e}
\end{equation}
where $\alpha$ is the attenuation constant $\alpha= \frac{\pi}{\lambda |\beta_g| Q}$ and 
$E_{d0}^{(i)}$ is the value of $E_{d0}$ at the beginning of the TW structure.\\
Assuming the structure is not so long, $\alpha L \ll 1$, and estimating $Z_e^{(SW)} \approx Z_e^{(TW)}$, 
from (\ref{6.3.1e}) - (\ref{6.3.3e}) we can estimate the ratio:
\begin{equation}
\frac{V_d^{(TW)}}{V_d^{(SW)}} \approx \sqrt{\frac{2 \pi L }{\lambda |\beta_g| Q}}.
\label{6.3.4e}
\end{equation} 
For structures operated as TW the choice of $|\beta_g|$ is of primary importance, simultaneously 
defining both positive and negative DS properties - RF efficiency (\ref{6.3.2e}) as 
$\sim (|\beta_g|)^{-\frac{1}{2}}$, wave attenuation $ \alpha \sim  (|\beta_g|)^{-1}$, 
phase distribution sensitivity  to cell frequencies deviations $\sim (|\beta_g|)^{-1}$.
A typical value of $(|\beta_g|) \sim 10^{-2}$ is usually accepted as a compromise.\\
For $S$-band applications, $\lambda = 0.1 m, Q \sim 10^4$ one can conclude from (\ref{6.3.4e}) -
for structure length L $<$ 1 m SW operating regime is more effective to obtain required value of
$V_d$.
\section{Parameters of the DLW structure}
\begin{figure}[htb]
\centering
\epsfig{file=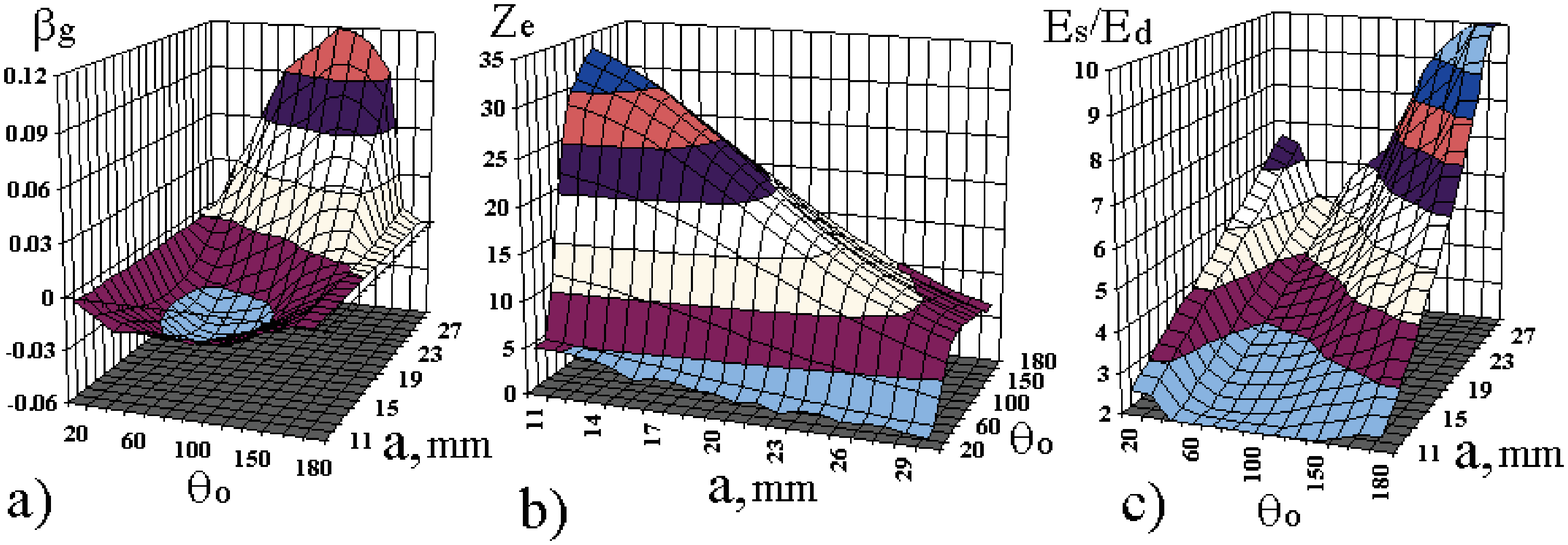, width =140.0mm}
\begin{center}
Figure 6: The surfaces $\beta_g(a,\Theta_0)$ (a), $Z_e(a,\Theta_0)$ (b) and $\frac{E_s}{E_d}(a,\Theta_0)$ for a DLW structure, $\frac{t_d}{\lambda}=0.054, \lambda = 10 cm$.
\end{center}
\label{6f}
\end{figure}
\begin{figure}[htb] 
\centering
\epsfig{file=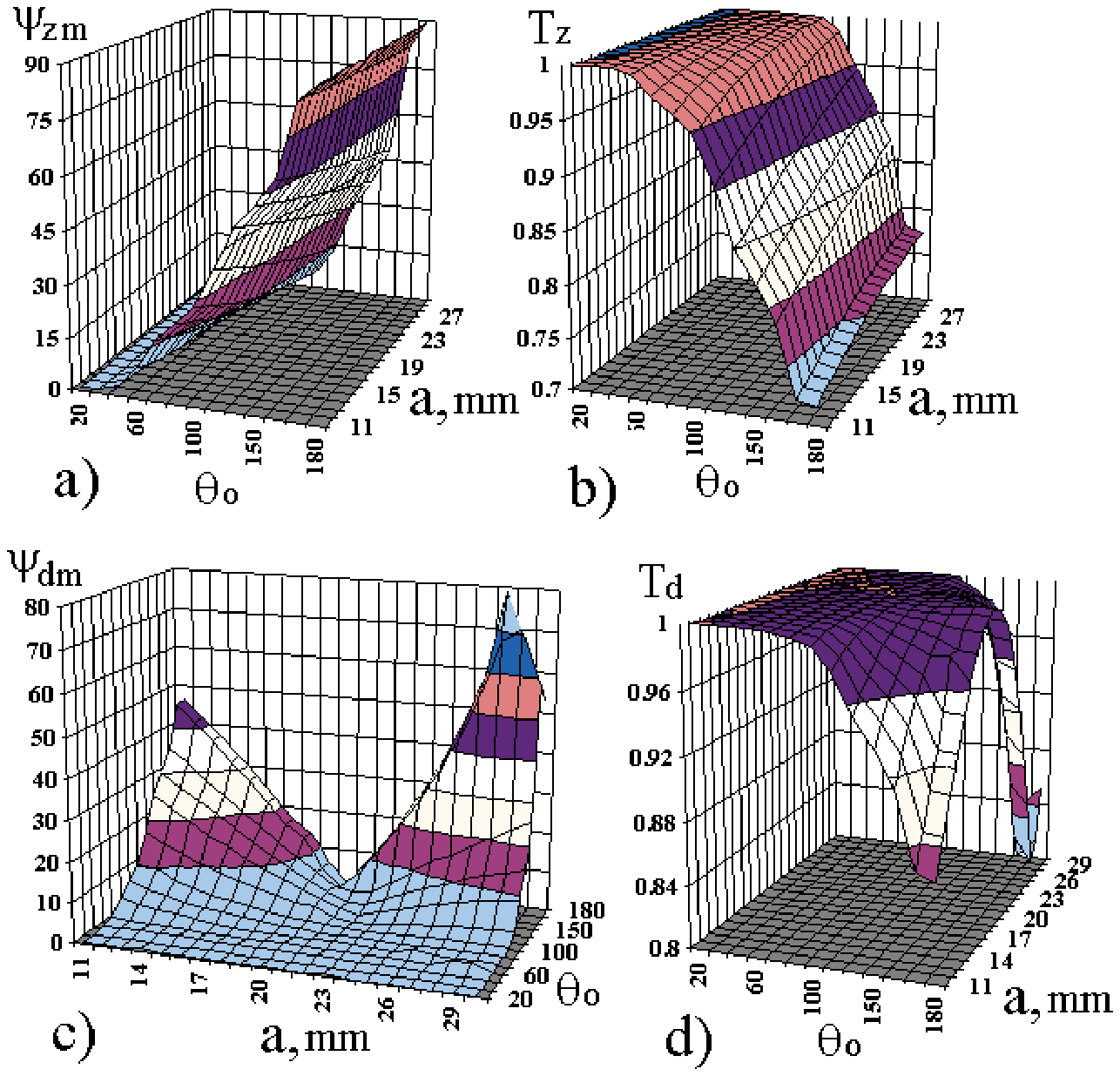, width =140.0mm,height=90.0 mm}
\begin{center}
 Figure 7: The surfaces $\Psi_{zm}(a,\Theta_0)$ (a), $T_z(a,\Theta_0)$ (b), $\Psi_{dm}(a,\Theta_0)$ (c) and 
$T_d(a,\Theta_0)$ (d) for a DLW structure, 
$\frac{t_d}{\lambda}=0.054, \lambda = 10 cm$.
\end{center}
\label{7f}
\end{figure} 
The study of DLW parameters has been performed in a wide range of operating parameters $\frac{\pi}{9} \leq 
\Theta_0 \leq \pi$ and for aperture radii $a$ in the range of $ 0.11 \leq \frac{a}{\lambda} \leq 0.30 $, 
assuming $\lambda =10 cm$. Always the $E_z$ distribution is calculated along the line $x=2 mm$. The 
results are presented in Fig. 6 and Fig. 7 as two dimensional surfaces to have general 
view for the DLW properties. In Fig. 6 the surfaces for group velocity $\beta_g(a,\Theta_0)$, Fig. 6a, 
effective transverse shunt impedance $Z_e(a,\Theta_0)$ and ratio $\frac{E_s}{E_d}(a,\Theta_0)$, 
where $E_s$ is the maximal electric field at the surface, are plotted.\\  
As it is common for slow wave structures, $\beta_g(a,\Theta_0) \rightarrow 0, \Theta_0 \rightarrow 
0, \pi$. For each intermediate value of $\Theta_0 \neq  0, \pi$ there is the inversion point, 
(\ref{2.5.2e}),  $(\frac{a}{\lambda})_{inv}(\Theta_0)$ corresponding to $\beta_g=0$. Below this point  
$(\frac{a}{\lambda}) < (\frac{a}{\lambda})_{inv}$ the group velocity is negative 
$\beta_g(\frac{a}{\lambda}) <0 $, corresponding to the backward traveling wave. For  
$(\frac{a}{\lambda}) > (\frac{a}{\lambda})_{inv}$ the group velocity is positive $\beta_g(\frac{a}{\lambda}) >0$,
corresponding to the forward wave. \\ 
The surface $Z_e(a,\Theta_0)$ has an upland for $\Theta_0 \approx \frac{\pi}{2}$ and the general trend is that
$Z_e$ decreases with increasing $a$.\\
The surface of the ratio $\frac{E_s}{E_d}$ has a valley for $\Theta_0 \approx \frac{\pi}{3}$ and the general 
trend is that $\frac{E_s}{E_d}$ increases with increasing $a$.\\
In Fig. 7 the surfaces $\Psi_{zm}(a,\Theta_0)$, Fig. 7a, $T_z(a,\Theta_0)$, Fig. 7b, 
$\Psi_{dm}(a,\Theta_0)$, Fig. 7c, and $T_d(a,\Theta_0)$, Fig. 7d, which are related also to the 
quality of the field distributions, (\ref{4.3e}) are shown.\\  
The transit time parameter is a more usual value in the list of features for periodical structures 
and describes also the relative weight of the synchronous harmonic in the total distribution. 
Always there is a connection – a smaller value of $\Psi_j$ leads to a higher value of $T_j$.\\  
For the original field component $E_z$ the maximal phase deviation $\Psi_{zm}(a,\Theta_0)$ rises
fast with $\Theta_0$, see Fig. 5a, and for $\Theta_0 \rightarrow \pi$ (SW mode)  $\Psi_{zm}(a,\Theta_0)
\rightarrow \frac{\pi}{2}$. It reflects the natural attenuation (\ref{4.1.4e}) for higher spatial harmonics 
at low values of $\Theta_0$. The corresponding 
$T_z(a,\Theta_0)$ reduction can be seen in Fig. 7b and  for $\Theta_0 \rightarrow \pi$, \quad $T_{z}(a,\pi)
\rightarrow \approx 0.75$. For larger values of aperture radii $a$ $\Psi_{zm}$ rises slower, also due to the stronger 
attenuation of higher spatial harmonics from the iris $r=a$ to the axis $r=0$.\\
For small aperture radii $(\frac{a}{\lambda}) < (\frac{a}{\lambda})_{inv}$ the components 
of the deflecting field $e_{r0}, h_{\vartheta 0} $ in a DLW have opposite phasing and the deflecting 
field $E_d$ reaches smaller values of $\Psi_{dm}(a,\Theta_0)$, as compared to the original field components, 
Fig. 7c. It reflects the reduction of $e_{dp}$ amplitudes, according to (\ref{3.2.6e}) for the opposite 
phasing of $HE_1$ and $HM_1$ waves. For small values of $\Theta_0$ $\Psi_{dm}(a,\Theta_0) < \Psi_{Exm, Hym}(a,\Theta_0)$ 
because the corresponding values for contributing components are small too and additionally the
compensation due to opposite $E_x, H_y$ phasing works. This compensation essentially decelerates the rising of 
$\Psi_{dm}(a,\Theta_0)$ with increasing $\Theta_0$.\\
With $\Theta_0 \rightarrow \pi$ the TW mode tends to the SW with $\Theta_0 = \pi$. But the distribution of $E_d$ for a SW mode 
has the properties of an 'equivalent' traveling wave, (\ref{6.1.5e}). 
For the reduction of $\Psi_{dm}(a,\Theta_0)$ both opposite phasing and the balance of the amplitudes $e_{x0}, h_{y0}$ are essential.
The second component becomes important for $\Theta_0 \approx \pi$. The balance between the amplitudes 
$e_{x0}$ and $h_{y0}$ depends on the aperture radius $a$, (\ref{2.5.3e}) and for 
$e_{x0} \sim - Z_0 h_{y0}$ we get $\Psi_{dm}(a,\pi) \sim 0$ instead of 
$\Psi_{E_x} = \Psi_{H_y}=\frac{\pi}{2}$. One can see a clear canyon in the $\Psi_{dm}(a,\Theta_0)$ surface 
in Fig. 7c and a corresponding arc in the $T_{d}(a,\Theta_0)$ surface in Fig. 7d for $\Theta_0 \rightarrow \pi$.
Unfortunately, the bottom of this canyon is not so far from the curve $(\frac{a}{\lambda})_{inv}(\Theta_0)$ 
and not so close to the region with high values of $Z_e$.
\subsection{Parameters for TW mode}
\begin{figure}[htb]
\centering
\epsfig{file=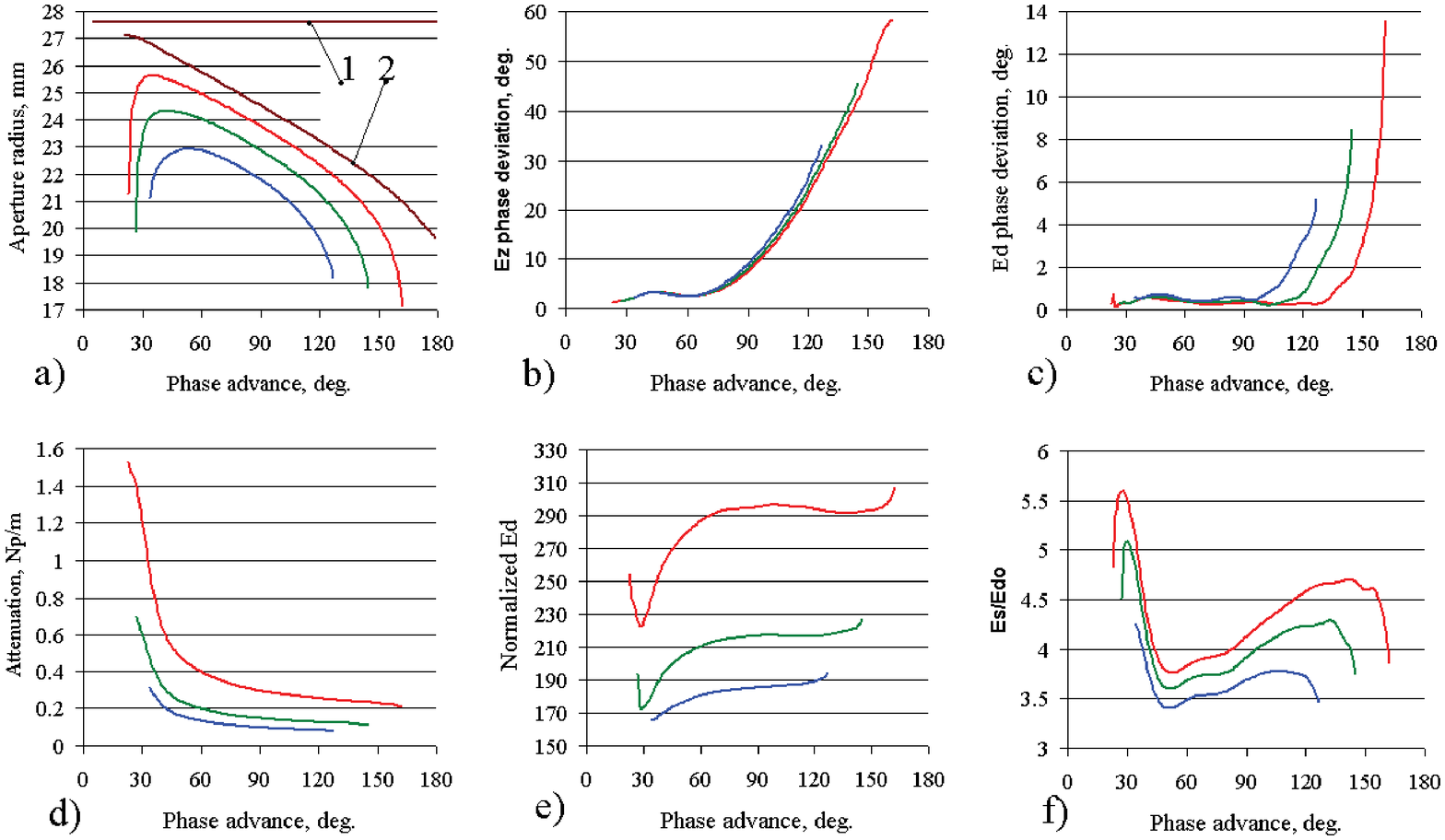, width =140.0mm,height = 90.0mm}
\begin{center}
 Figure 8: The dependences on $\Theta_0$ for aperture radius $a$ (a), the maximal phase deviation 
$\Psi_{zm}$ (b), $\Psi_{dm}$ (c), attenuation $\alpha$ (d), normalized field 
$\frac{E_{d0} \lambda}{\sqrt{P_t}}$ (e) and ratio $\frac{E_s}{E_d}$ (f) for relative 
group velocity values $\beta_g=-0.01$ (red curves), $\beta_g=-0.02$ (green curves) and 
$\beta_g=-0.03$ (blue curves).
\end{center}
\label{8f}
\end{figure}        
In more details the DLW parameters for the TW mode are plotted in Fig. 8. The important TW parameter 
is $\beta_g$. In Fig. 8a the dependences of the aperture radius $a$ 
on the operating phase advance $\Theta_0$ is plotted as to get the required value of
$\beta_g$. The brown curve (2 in Fig. 8a) 
shows the calculated position of the inversion point $\beta_g=0$ in dependence on $\Theta_0$. According to the small pitch
 approximation, the 
inversion radius should be constant for all values of $\Theta_0$ and $a_{inv}=\frac{\sqrt{3}}{2 \pi} \lambda =27.57 mm$, 
(\ref{2.5.2e}). It is shown by the line 1 in Fig. 8a. The difference between the line 1 and the curve 2 points out the usability 
 for the small pitch approximation.\\
Also in Fig. 8a are plotted the required values of $a$ to get $\beta_g=-0.01$ (red curve), $\beta_g=-0.02$ (green curve) and 
$\beta_g=-0.03$ (blue curve). Other DLW parameters are plotted in Fig. 8b, f for the obtained values of $a$ 
with the respective colors.\\
The phase deviation $\Psi_{zm}$ (Fig. 8b) rises fast for $\Theta_0 > \frac{\pi}{2}$. 
Comparing plots in Fig. 8b and Fig. 8a, one can conclude, that for the condition $\beta_g=const$ 
 lower values of $\Theta_0$ are more important for an efficient attenuation of  higher harmonics 
(\ref{4.1.4e}) in the original field components.\\
Due to opposite phasing of the hybrid waves, the phase deviation $\Psi_{dm}$ for the deflecting field (Fig. 8c) 
rises not so fast and a low aberration level in the $E_d$ distribution is possible for $\Theta_0 \leq 
\frac{2 \pi}{3}$.\\ 
For small values of $\Theta_0 < \frac{\pi}{3}$ the DLW structure is closer to the small pitch approximation. 
The sequence of this approximation is a lower quality factor $Q$ and a higher attenuation constant 
$\alpha$ values, Fig. 8d. The total deflecting voltage $V_d$ (\ref{6.1.3e}) depends on both normalized 
field value $\frac{E_{d0} \lambda}{\sqrt{P_t}}$, Fig. 8e, and $\alpha$. Operation with $\Theta_0 < \frac{\pi}{3}$ 
should be considered additionally from the point of RF efficiency.\\
For the electrical strength of the structure the range $45^o \leq \Theta_0 \leq \frac{2\pi}{3}$ is preferable,
Fig. 8f.
\subsection{Parameters for SW mode}
Parameters of the DLW structure for SW mode are plotted in Fig. 9.\\
\begin{figure}[htb]
\centering
\epsfig{file=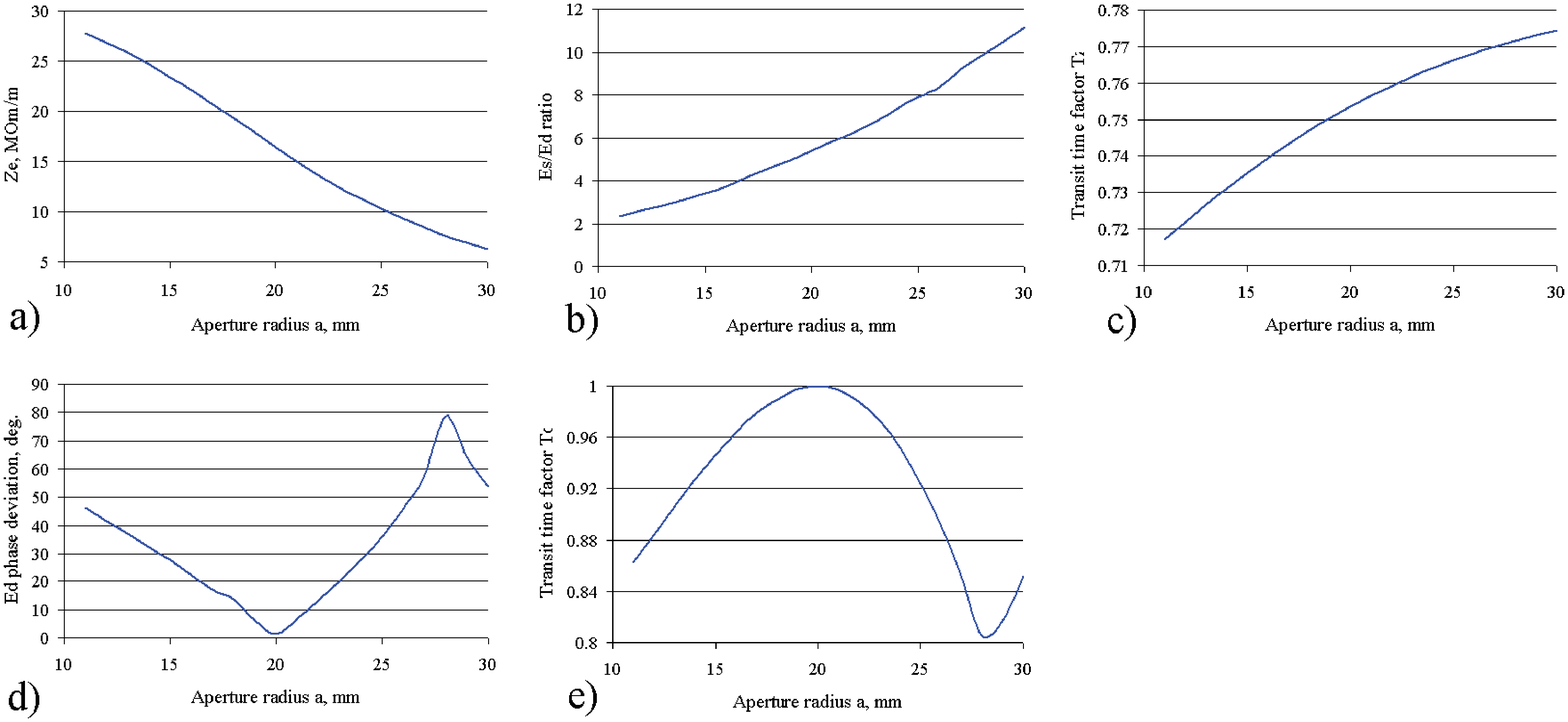, width =140.0mm}
\begin{center}
 Figure 9: The parameters of the DLW structure for SW mode in dependence on aperture radius.
Effective transverse shunt impedance $Z_e$ (a), ratio $\frac{E_s}{E_d}$ (b), transit time factor 
$T_z$ (c), the maximal phase deviation $\Psi_{dm}$ (d) and transit time factor $T_d$ (e) for the
deflecting field. 
\end{center}
\label{9f}
\end{figure}  
The effective shunt impedance $Z_e$ decreases with increasing $a$, Fig.9a, and the maximal electric field at the surface increases, Fig. 9c. 
The maximal phase deviation $\Psi_{zm}$ for the longitudinal 
$E_z$ component is always $\Psi_{zm}=\frac{\pi}{2}$, due to the step-wise rise in the $E_z$ phase in SW mode 
and $T_z$ is always $T_z \simeq 0.75$, Fig. 9c.\\
For the deflecting field $E_d$ the value of the phase deviation $\Psi_{dm}$ 
depends on the amplitude balance of $e_{x0}$ and $h_{y0}$, Fig. 9d, which defines also the value of $T_d \sim 1$, Fig. 9e. 
We can get $\Psi_{dm}  \approx 0$, see Fig. 9d, at some value of $a_0$, but it is close 
to the position of the inversion point. 
For DLW in SW mode with $\Psi_{dm} \approx 0$ exist limitations from the 
mode mixing problem as described below\\
As it was shown in Section 4.3, the SW operating mode is more efficient for obtaining a required value of $V_d$ 
with a short DS. As for the homogeneity of the longitudinal component $E_z$, essential reductions of aberrations 
(reduction of higher spatial harmonics) can not be realized for SW mode. As for aberrations in the transverse field $E_d$, we have a contradictory 
choice for DLW - either RF effective operation with smaller $E_s$ values, but significant 
level of nonlinear additions,
or operation with reduced aberrations at the expense of RF power and at higher surface field.
\section{Parameters of the decoupled TE-structure}
In DLW structures the aperture radius $a$ defines simultaneously both RF efficiency, and dispersion 
properties. In TE-structures (Fig. 1d) these parameters are separated. The RF efficiency depends 
mainly on the distance between the noses (effective aperture radius) and the disk thickness. The passband width can 
be changed by the inner disk radius $r_w$. 
This structure has no rotational symmetry. For the complete description of the field distribution, 
starting from (\ref{2.2e}), we have to add waves and harmonics with $n=3,5,7,...$ variations along 
azimuth, see Section 3.3. Below we do however not consider multipole additives. The nose tip shape was 
optimized to reduce the multipole additives and this shape was used for consideration.\\
The study of the RF parameters has been performed in a range of $\frac{\pi}{9} \leq \Theta_0 \leq \pi$ and 
$ 0.06 \leq \frac{a}{\lambda} \leq 0.15 $, assuming $\lambda =10 cm$ with all other parameters - the disk thickness $t_d  \approx 0.52 d$ and iris radius 
$\frac{r_w}{\lambda} = 0.24$ fixed from the previous RF efficiency optimization.\\
\begin{figure}[htb]
\centering
\epsfig{file=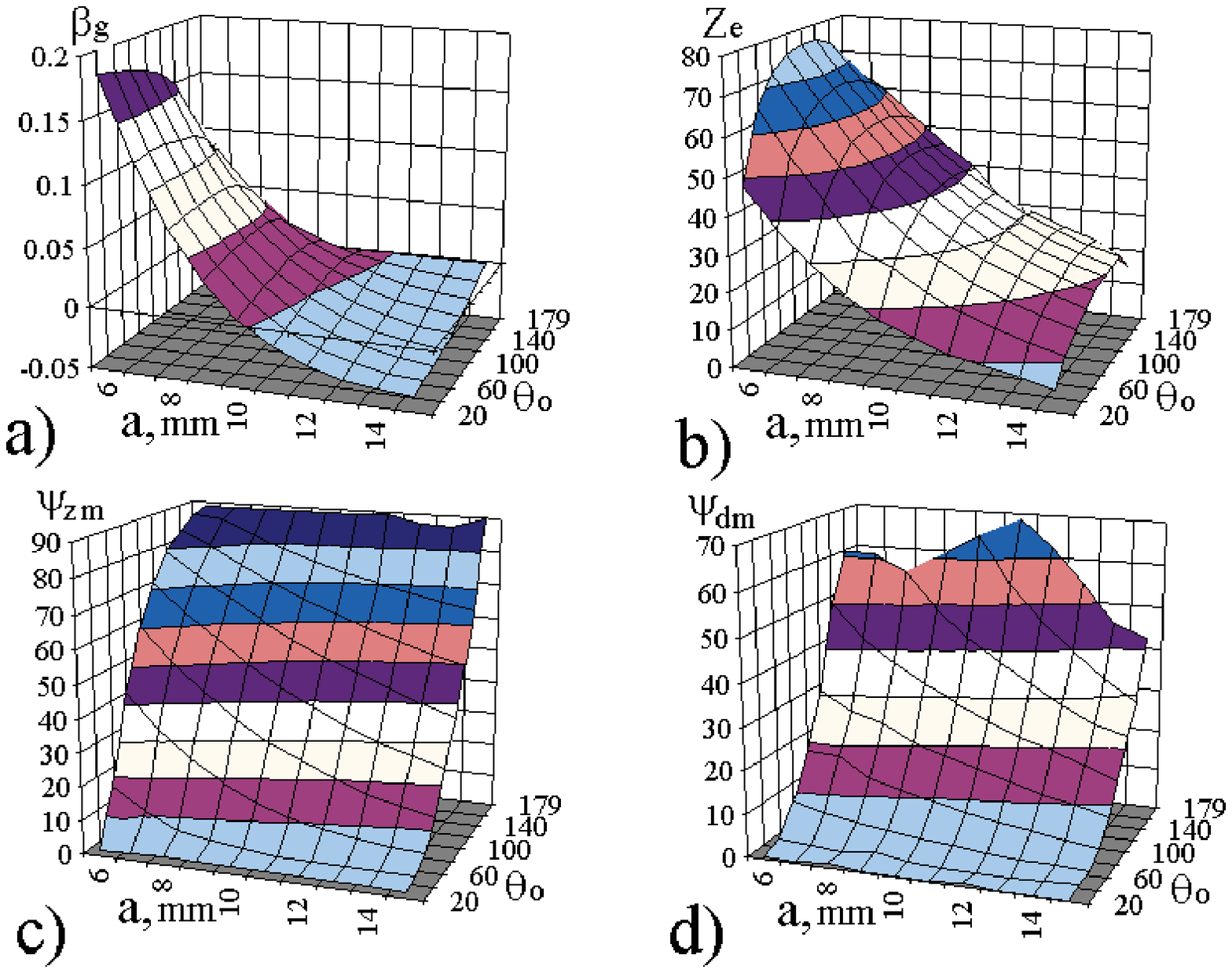, width =110.0mm,height= 90.0mm}
\begin{center}
Figure 10: The surfaces of $\beta_g(a,\Theta_0)$ (a), $Z_e(a,\Theta_0)$ (b) $\Psi_{zm}(a,\Theta_0)$ 
(c) and $\Psi_{dm}(a,\Theta_0)$ for the TE-structure, $\lambda = 10 cm$.
\end{center}
\label{10f}
\end{figure}
Results are presented in Fig. 10 as two dimensional surfaces. The range of aperture radii, considered for the
TE-structure and DLW, have a common region $ 0.11 \leq \frac{a}{\lambda} \leq 0.15 $. It allows us to compare the general 
behavior of structure parameters both qualitatively and quantitatively. 
For small aperture radii $\frac{a}{\lambda} \leq 0.11$ the structure has equal phasing of the synchronous harmonics 
$e_{x0}, h_{y 0} $ and, as a consequence, a positive group velocity $\beta_g >0 $, Fig. 10a, which decreases with aperture 
increasing radius. Also for each value of $\Theta_0$ we find the inversion point $(\frac{a}{\lambda})_{inv}(\Theta_0) \approx 0.11 $, 
corresponding to $\beta_g=0$. Above the inversion point $\beta_g <0$ and with further increasing $\frac{a}{\lambda}$ 
the TE-structure degenerates into a DLW.\\ 
For small aperture radii the TE-structure reaches high values of $Z_e$, Fig. 10b, which decrease with increasing $a$. 
For the same aperture radius both TE-structure and DLW have approximately equal values of $Z_e$, but differ strongly 
in the width of the passband.\\ 
Similar to a DLW, see Fig. 7a, the phase deviation $\Psi_{zm}(a,\Theta_0)$ rises with $\Theta_0$, 
Fig. 10c, but the rise is faster - for smaller values of $a$ the attenuation of higher harmonics in TE-structures, 
according to (\ref{4.1.4e}), is not so strong.\\ 
For equal $e_{x0}, h_{y 0} $ phasing the phase deviations $\Psi_{dm}(a,\Theta_0)$ for the deflecting field $E_d$ 
is larger than for the $E_x, H_y$ components and rises faster with $\Theta_0$ for small values of $a$. According to (\ref{3.2.6e}), 
the amplitudes of the spatial harmonics in the $E_d$ description, are enlarged for equal phasing of $HE_1$ and $HM_1$ waves as 
compared to opposite phasing.\\
For values of $a$ above the inversion point the harmonics $e_{x0}$ and $h_{y0}$ obey an opposite phasing and the partial 
compensation starts, leading to a reduction of $\Psi_{dm}(a,\Theta_0)$, which is however not very effective. For the compensation of the  
harmonics both opposite phasing and comparable values 
for amplitudes are essential. In the TE-structure with introduced noses, for the considered range of parameters, 
the amplitude of $e_{x0}$ dominates with respect to the amplitude of $Z_0 h_{y0}$ and the phase deviation $\Psi_{dm}(a,\Theta_0)$ for the $E_d$
is defined mainly by $\Psi_{xm}(a,\Theta_0)$ for the $E_x$ component, Fig. 10d, which rises with $\Theta_0$ as all 
original field components. In the considered range of parameters this is the reason for the absence of a canyon in the dependence 
of $\Psi_{dm}(a,\Theta_0)$ on $a$ for $\Theta_0 \sim 180^0$, as we have seen for the DLW structure in  Fig. 7c, Fig. 9d.
\subsection{Parameters for TW mode}
\begin{figure}[htb]
\centering
\epsfig{file=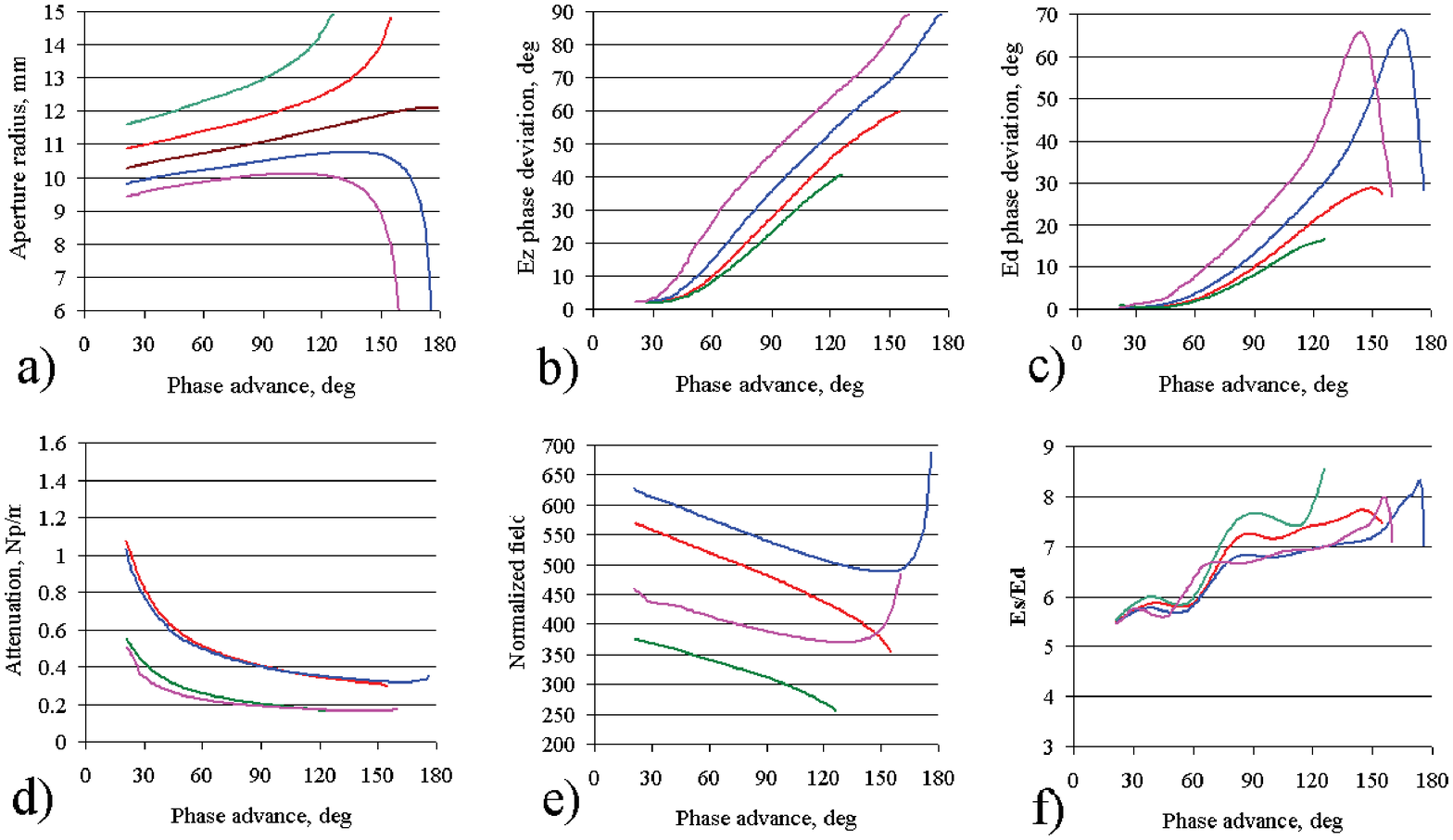, width =150.0mm}
\begin{center}
 Figure 11: The dependences on $\Theta_0$ on the aperture radius $a$ (a), the maximal phase deviation 
$\Psi_{zm}$ (b), $\Psi_{dm}$ (c), attenuation $\alpha$ (d), normalized field 
$\frac{E_{d0} \lambda}{\sqrt{P_t}}$ (e) and ratio $\frac{E_s}{E_d}$ (f) for relative 
group velocity values of $\beta_g=-0.01$ (red curves), $\beta_g=0.01$ (blue curves), 
$\beta_g=-0.02$ (green curves) and $\beta_g=0.02$ (magenta curves).
\end{center}
\label{11f}
\end{figure} 
Regarding RF efficiency in the TW mode, small values of $a$ are not attractive, because they
result in high values of $\beta_g$ and related lower values of normalized field $\frac{E_{d0} \lambda}{\sqrt{P_t}}$. 
The inversion of the sign of $\beta_g$ allows, similar to the DLW, a reasonable choice of $\beta_g$. The dependence 
of the inversion value $(\frac{a}{\lambda})_{inv}$ on $\Theta_0$ is shown in Fig. 11a with a brown curve.\\
Parameters of the TE-structure for TW mode are plotted in Fig. 11 in the same style with equivalent 
parameters as for the DLW in Fig. 8. Comparing the plots in Fig. 10b, c and Fig. 8b, c, one can see for the same value of
$\beta_g$ larger phase deviations $\Psi_{zm}(a,\Theta_0)$ for $E_z$, Fig. 11b, Fig. 8b, and essentially larger 
phase deviations $\Psi_{dm}(a,\Theta_0)$ for $E_d$, Fig. 11c, Fig. 8c in the TE-structure. These results are the consequences of small 
values of $a$ and equal phasing of $e_{x0} and h_{y0}$.\\ 
As compared to a DLW, the structure has a lower $Q$-factor, resulting in a higher attenuation constant $\alpha$, as one 
can see by comparing the plots in Fig. 11d and Fig. 8d. But the normalized field $\frac{E_{d0} \lambda}{\sqrt{P_t}}$ (\ref{6.3.2e}) is 
essentially higher, see plots in Fig. 11e and 
Fig. 8e. The total deflecting voltage $V_d$ depends on the normalized field, the attenuation constant and the structure 
length $L$ (\ref{6.3.3e}). Instead of a higher $\alpha$, the reserve in the value of $\frac{E_{d0} \lambda}{\sqrt{P_t}}$ is 
sufficient to find values of $L$ with higher RF efficiency. As one can see from Fig. 11e, f, regarding RF efficiency a TW 
operation with $\beta_g > 0$ is preferable - as compared to $\beta_g <0$ a higher value $\frac{E_{d0} \lambda}{\sqrt{P_t}}$ is reached
for the same $\alpha$. But the phase deviations, both $\Psi_{zm}$ and $\Psi_{dm}$, are larger for 
$\beta_g >0$, see Fig. 11b, c. \\
Due to the noses and, especially, the nose end shape, TE-structures reach higher $\frac{E_s}{E_d}$ ratios, 
see Fig. 11e and Fig. 8e. Generally, DS's with equal $E_x$ and $H_y$ phasing should have higher values of the ratio $\frac{E_s}{E_d}$.
The maximal electric field at the surface $E_s$ is connected with $E_x$. For the opposite phasing of the $e_{x0} and h_{y 0} $ 
components $E_x$ and $H_y$ contribute together, (\ref{3.2e}) to the creation of $E_d$. For equal $e_{x0}, h_{y 0} $ phasing the
magnetic field partially compensates the deflection from the electric field and a higher value of $E_x$ is 
required to produce the same deflecting effect.
\subsection{Parameters for SW mode}
\begin{figure}[htb]
\centering
\epsfig{file=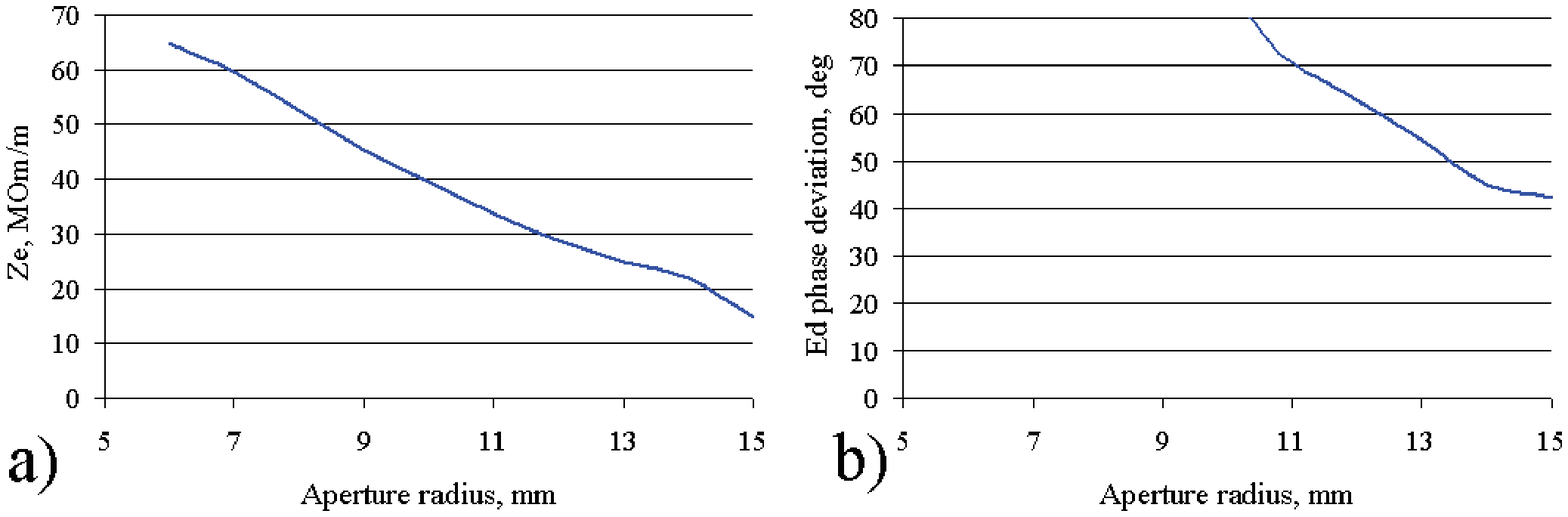, width =120.0mm}
\begin{center}
 Figure 12: The shunt impedance $Z_e$ (a) and the maximal phase deviation $\Psi_{dm}$ (b) for SW mode in dependence 
on the aperture radius $a$. 
\end{center}
\label{12f}
\end{figure}
Due to higher values of $Z_e$, Fig. 12a, SW mode was originally proposed as preferable application of TE-structures.
As compared to a DLW, a decoupled TE-structure has additional freedom in the dimensions - nose opening along azimuth and inner disk radius, 
which differs from the aperture radius. The cell dimensions were optimized for RF efficiency assuming 
an $L$-band operation with $a=15.5 mm$, corresponding to $a=6.72 mm$ in the $S$-band range.\\  
The phase deviation $\Psi_{zm}$ shows a typical SW mode behavior at $\Psi_{zm}=\frac{\pi}{2}$. 
Concerning the deflecting field $E_d$ the dependence of $\Psi_{dm}$ on $a$ is more interesting, Fig. 12c. In the 
range $5 mm < a < 11 mm$ (below the inversion point), hybrid waves have equal  phasing; the 'equivalent' traveling wave in (\ref{6.1.5e}) is the backward wave
with respect to the particle motion and $\Psi_{dm} \approx \frac{\pi}{2}$.\\ 
 Above the inversion point $a > 11 mm$ we find a negative structure dispersion, opposite 
$e_{x0}, h_{y 0} $ phasing and $\Psi_{dm}$ decreases with further increasing $a$. If the plot of 
$\Psi_{dm}$ in Fig. 12b would be continued to $a \approx 20 mm$, a second inversion point and 
a valley in the plot of $\Psi_{dm}$ would appear, similar to the valley in the plot of $\Psi_{dm}$ for the DLW in Fig. 9d, because the
TE-structure degenerates in its properties to a DLW for large values of $a$.\\
The decoupled TE-structure has been considered here mainly in contrast to a DLW. For a small aperture 
radius this DS shows an equal $e_{x0}, h_{y 0} $ phasing, a high RF efficiency, but a poor deflecting 
field quality for bunch rotation. In the DLW consideration the necessity of opposite $e_{x0}, h_{y 0} $ phasing 
was illustrated to achieve a high quality of the deflecting field, but at the expense of RF efficiency.\\
The main value of this study is the detection of the dispersion inversion at a small aperture radius. 
After that the advantage of the decoupled control over the field distribution was used to combine RF efficiency 
and deflecting field quality, \cite{defl_min}.
\section{Hybrid waves phasing and balance.}
In the following plots for field distributions are shown to compare the effect of both the opposite 
and the equal phasing of $e_{x0}$ and $h_{y0}$.\\
A DS with a total length of $200 mm = 2.0 \cdot \lambda$ is considered for operation in 
TW and SW mode with the same average deflecting field of $E_{d0} = 1 MV/m$. To emphasize the particularities 
in distributions, some frames in plots have different scales of the field strength.\\ 
\begin{figure}[htb]
\centering
\epsfig{file=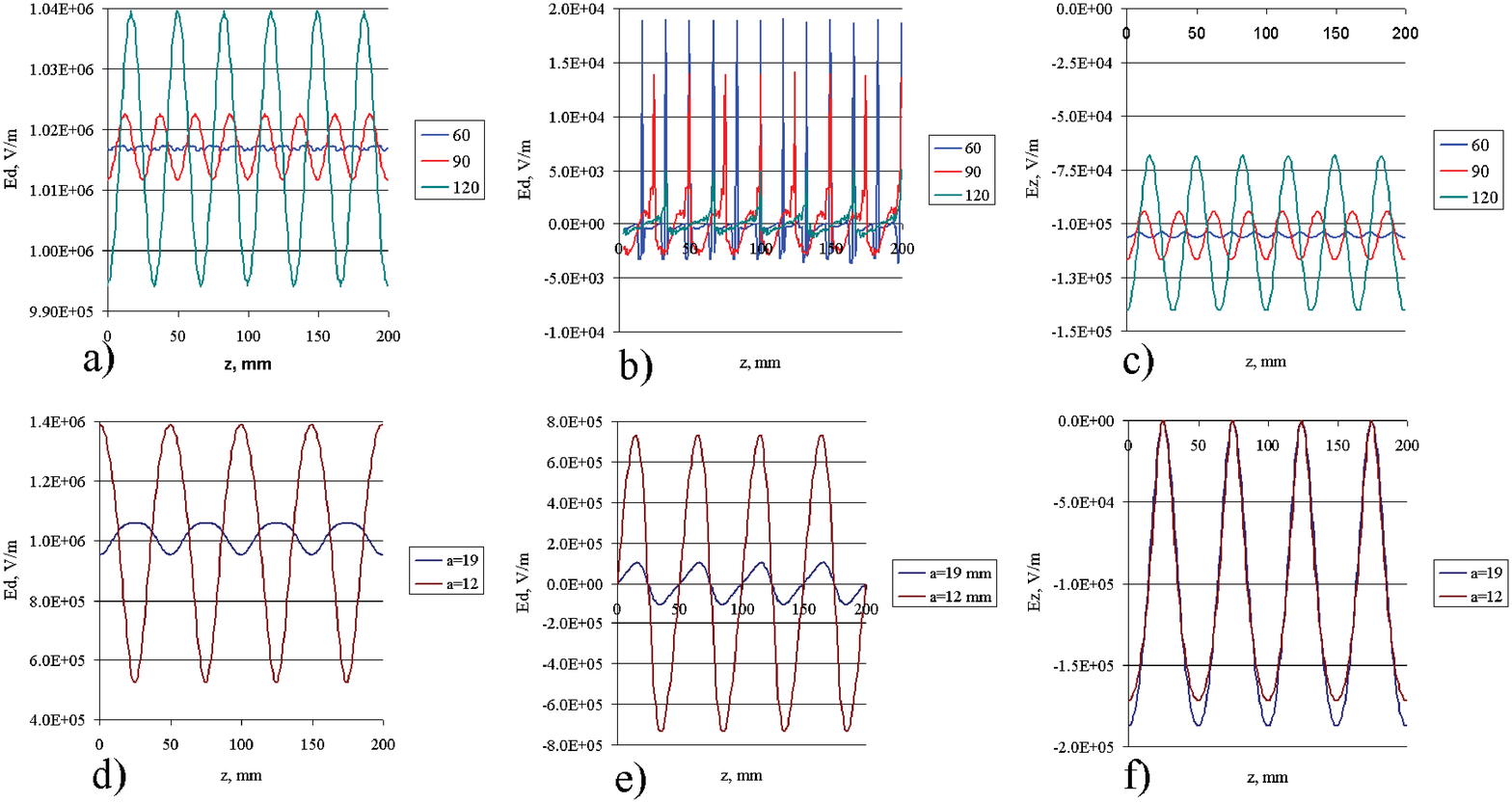, width =140.0mm,height=90.0 mm}
\begin{center}
 Figure 13: Plots of $E_d and E_z$ distributions for DLW TW (upper row) and SW mode 
(bottom row). TW mode with $\Theta_0=\frac{\pi}{3}, \frac{\pi}{2}, \frac{2\pi}{3}$  - blue, red and green curves, respectively. 
SW mode for $\frac{a}{\lambda}=0.12, 0.194, $ brown and dark blue curves, respectively.
The plotted distributions show $E_d , \phi=0$, (a), (d) - $E_d, \phi=\frac{\pi}{2}$, (b), (e)
and  $E_z, \phi=\frac{\pi}{2}$ (c), (f).
\end{center}
\label{13f}
\end{figure}
In Fig. 13 plots of distributions for opposite $e_{x0}, h_{y0}$ phasing are shown, for a DLW structure with large aperture 
radius $\frac{a}{\lambda} \sim 0.23$. A detailed description of the plots is given in the caption 
of Fig. 13. For direct comparison equivalent plots are shown in Fig. 14 for the decoupled TE structure 
with equal $e_{x0}, h_{y0}$ phasing and small aperture radius $\frac{a}{\lambda} \sim (0.07 - 0.11)$.
In addition to plots for equal phasing, corresponding to $\beta_g > 0$, plots for opposite phasing, i.e. $\beta_g < 0$ are presented in Fig. 14.\
 Numerical data - amplitudes of the spatial harmonics, 
calculated according (\ref{4.1.2e}) - are presented in the Table 3.\\
\begin{figure}[htb]
\centering
\epsfig{file=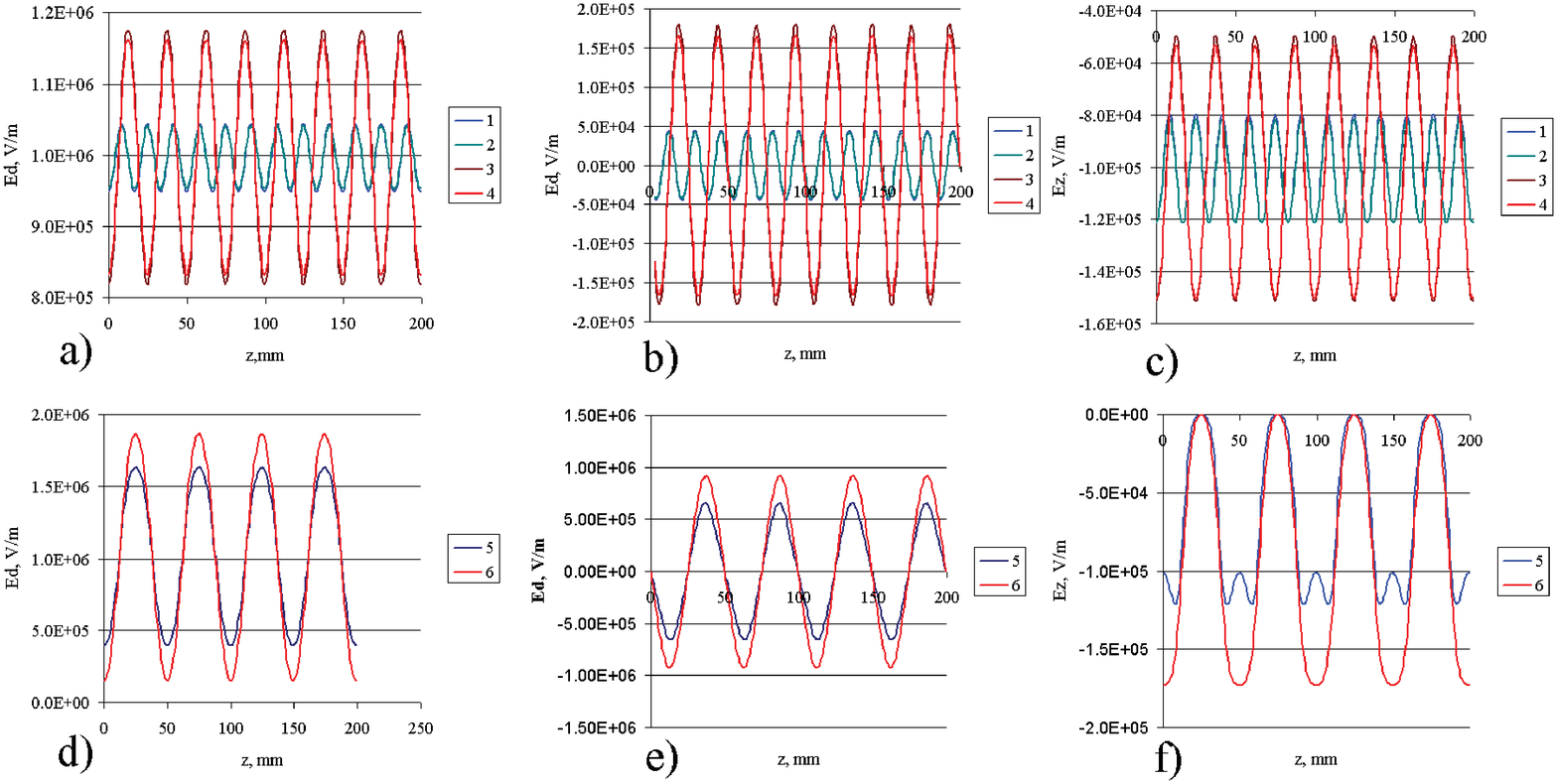, width =140.0mm,height=90.0 mm}
\begin{center}
 Figure 14: Plots of $E_d$ and $E_z$ distributions in the decoupled TE structure for TW 
 (upper row) and SW (bottom row) operation. TW mode with $\Theta_0=\frac{\pi}{3}, \beta_g=-0.01$ , 
$\Theta_0=\frac{\pi}{3}, \beta_g=0.01$, $\Theta_0=\frac{\pi}{3}, \beta_g=-0.02$  and $\Theta_0=\frac{\pi}{2}, \beta_g=0.02$ - 
blue, red, green and brown curves, respectively. SW mode for $\frac{a}{\lambda}=0.0672, 0.108$ -
dark blue and red curves, respectively. 
The plotted distributions show $E_d, \phi=0$, (a), (d) - $E_d, \phi=\frac{\pi}{2}$, (b), (e)
and  $E_z, \phi=\frac{\pi}{2}$ (c), (f).
\end{center}
\label{14f}
\end{figure}
\begin{table}[htb]   
\begin{center}
\centering{Table 3: The relative amplitudes for higher spatial harmonics in the field distributions 
for the DLW and decoupled TE structures.}
\begin{tabular}{|l|c|c|c|c|c|c|c|c|c|c|c|c|c|c|c|c|}
\hline
DS, Operation,     &$\Theta_0$ &$E_z$, &$E_z$, &$E_z$, & $E_d$,& $E_d$,& $E_d$,\\
    $a= , mm$      &           &$e{z1}$&$e_{z2}$&$e_{z3}$&$e_{d1}$&$e_{d2}$&$e_{d3}$   \\
\hline
DLW,TW,$23.80 mm$  & 60        &-0.0048&-0.0007&-0.0010&0.0006 &-0.0009& 0.0006 \\
DLW,TW,$23.58 mm$  & 90        &-0.0581&-0.0004&-0.0003&0.0009 &-0.0001&-0.0001 \\
DLW,TW,$22.33 mm$  & 120       &-0.1950&-0.0048&0.0001 &0.0020 &-0.0015& 0.0012 \\
\hline
DLW,SW,$12.0 mm$   & 180       &-0.6103&-0.1385&-0.0348&-0.3818&-0.0623&-0.0008 \\
DLW,SW,$19.4 mm$   & 180       &-0.5619&-0.0691&-0.0077&-0.0271&-0.0148& 0.0047 \\
\hline
TE-,TW,$10.45 mm$  & 60        &-0.1131&-0.0002& 0.0000&0.0227 &-0.0001& 0.0002 \\
TE-,TW,$11.39 mm$  & 90        &-0.2520&-0.0024&0.0006 &0.0896 & 0.0002&-0.0001 \\
\hline
TE-,SW,$6.72 mm$   & 180       &-0.5178& 0.0998&0.1165 &-0.3010&-0.0194& 0.0124 \\
TE-,SW,$10.08 mm$  & 180       &-0.4842& 0.0640& 0.0472&-0.4384&-0.0064& 0.0076 \\
\hline
\end{tabular}
\label{3t}
\end{center}
\end{table} 
Comparing the distributions of the original field component $E_z$ for TW mode, $\Theta_0=\frac{\pi}{3}$, in Fig. 13c, 
Fig. 14c and the $e_{zp}$ coefficients in the Table 3, we see the effect of a large aperture radius, 
which provides a stronger attenuation of harmonics toward the axis, (\ref{4.1.4e}). The amplitudes of higher 
harmonics in $e_{zp}$ are essentially smaller in DLW. With increasing $\Theta_0$ the amplitudes 
of harmonics in the original field components at the axis rise due to weaker attenuation.\\  
For the deflecting field $E_d$ we directly see from Fig. 13 a, b, Table 3, a reduction of harmonic amplitudes 
$|e_{dp}| \ll |e_{zp}|$, as compared to the original field components, for opposite 
$e_{x0}, h_{y0}$ phasing both for bunch deflection $\phi =0$ and for bunch rotation $\phi =\frac{\pi}{2}$.
The spikes in Fig. 13b ($E_d, \phi=\frac{\pi}{2}$) reflect just numerical noise in the simulation of the fields; the
regular effects in the deviations of $E_d$ are much below the noise. According to (\ref{3.2.6e}), for opposite 
phasing of $HE_1$ and $HM_1$ waves in the field description (\ref{2.3.3e}) the higher spatial harmonics 
in the $E_x$ and $H_y$ components compensate each other on the background of an increased total main harmonic.\\ 
For equal $e_{x0}, h_{y0}$ phasing we do not see, Fig. 14 a, b, Table 3, a reduction of the harmonics amplitudes 
in $E_d$ in comparison with the original field components.\\
The advantage of opposite $e_{x0}, h_{y0}$ phasing for the suppression of harmonics is especially brightly shown 
in the TW mode in Fig. 14 a, b, c - for the negative group velocity (opposite phasing) the 
deviation from the average (due to the higher spatial harmonics) is always larger, as compared 
to the same value but positive group velocity (equal phasing).\\ 
For a TW mode with low value of $\Theta_0 \leq \frac{\pi}{2}$ the amplitude balance $|\frac{Z_0h_{y0}}{e_{x0}}|$ 
is not of primary importance - for harmonics suppression in the deflecting field the attenuation 
(\ref{4.1.4e}) in original field components is more essential. The plots of the $\frac{h_{y0}}{e_{x0}}$ balance 
are given in Fig. 15a together with the corresponding plots of $\Psi_{dm}$ both for TW and SW mode. 
The black dotted curve in Fig. 15a shows the plot of the $\frac{Z_0h_{y0}}{e_{x0}}$ balance in a DLW according to the 
small pitch approximation, (\ref{2.5.3e}). As can be seen from Fig. 15b, for a DLW in TW mode with 
$\Theta_0 =\frac{\pi}{3}$ the phase deviations $\Psi_{dm}$ are very small for all values of the aperture radius 
$a$, even though the harmonic amplitudes ${Z_0h_{y0}}$ and ${e_{x0}}$ are not balanced, Fig. 15a.\\ 
\begin{figure}[htb]
\centering
\epsfig{file=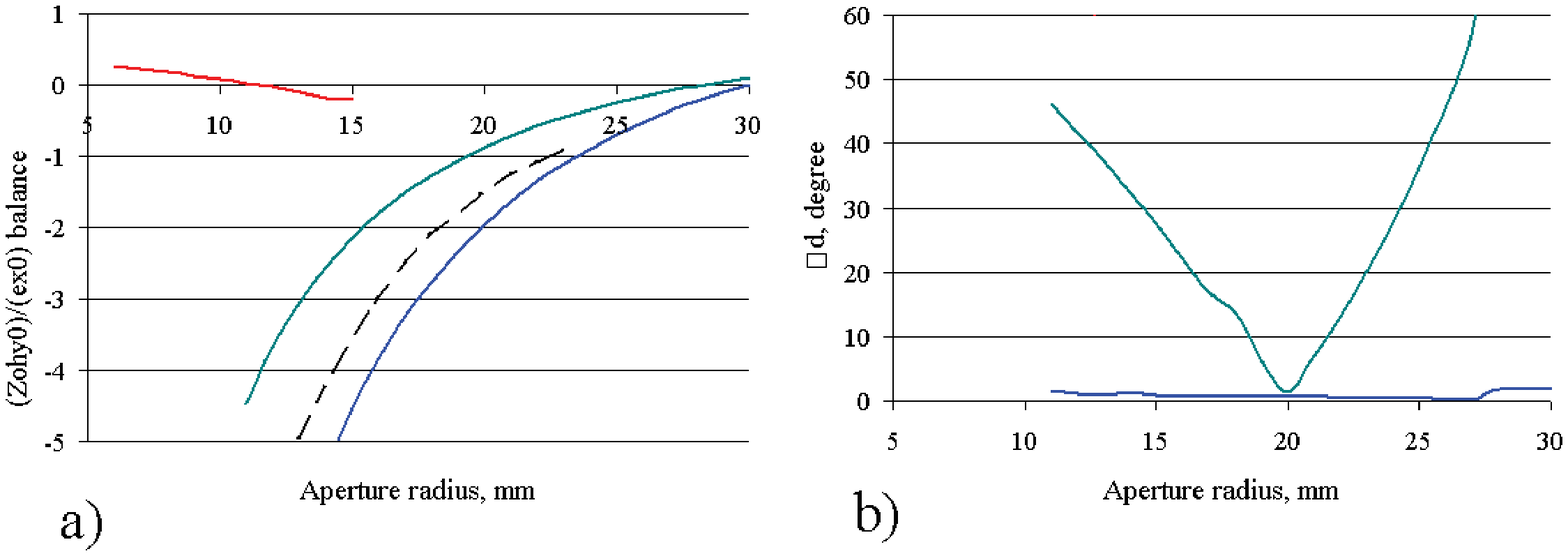, width =130.0mm}
\begin{center}
 Figure 15: Plots of $\frac{Z_0h_{y0}}{e_{x0}}$ balance (a) and corresponding plots of $\Psi_{dm}$ (b)
 for a DLW in TW mode ($\Theta=\frac{\pi}{3}$, blue curves) and SW mode, green curves. The red curve represents the balance of
$\frac{Z_0h_{y0}}{e_{x0}}$ for a decoupled TE-structure in SW mode.
\end{center}
\label{15f}
\end{figure}
The role of the balance of $\frac{Z_0h_{y0}}{e_{x0}}$ increases with $\Theta_0 \rightarrow \pi$ and for a SW 
operation it is the only way to suppress harmonics in a deflecting field. The aperture radius 
$\frac{a}{\lambda}=0.194$ for a DLW in SW mode corresponds to the minimal value of phase deviation of $\Psi_{dm} \approx 1.8^o$
 and $\frac{Z_0h_{y0}}{e_{x0}} \approx -0.85$, Fig. 15b,c. Comparing the amplitudes $e_{dp}$ in the Table 3 for $\frac{a}{\lambda}=0.12$ 
and $\frac{a}{\lambda}=0.194$, one can see a reduction of $e_{d1}$ by more than an order of magnitude and a reduction of $e_{d2}$ by 
more than a factor of four. But the deflecting field deviations are suppressed for $\frac{a}{\lambda}=0.194$ 
in 'total', i.e. on the background of the first harmonic $e_{d1}$ which is reduced by one order of magnitude we see in the distribution of the second harmonic 
$E_d, \phi=\frac{\pi}{2}$ in Fig. 13e $e_{d2}$ - $E_d$ non symmetric deviations.\\
Differing from the prediction of $e_{dxp}(x,y) \approx 0$ for $A=-B$, (\ref{3.2.6e}), the results of the numerical 
simulations show for a SW mode a minimal value $\Psi_{dm} \approx 2^o$ at $B \approx -0.87 \cdot A$. 
In this case we realize a DS option with damped $e_{d p}^{o}$ terms in (53), when the not compensated 
contribution of $e_{r0} + Z_0 h_{\vartheta 0}$ to the oscillations in (52) from the synchronous harmonics in $E_r$ and $H_{\vartheta}$ is damped 
by the first harmonics $e_{r1} +  Z_0 h_{\vartheta 1}$ and simultaneously the amplitudes for higher 
harmonics in $E_d$ are reduced, according to (\ref{3.2.6e}). 
\subsection{Structures classification}
The classification of a structure with a complicated field distribution is always rather conditional. 
Often the visual characteristics of the structure or field distribution are used. For example, the DS with 
evidently dominating transverse electric field, Fig. 1d, was treated in \cite{te_defl} as TE-type. In Fig. 16 it 
is shown, how this DS can be continuously transformed, with an appropriate transformation of the field distribution, 
into other DS's.\\
Suppose the starting point is (b) in Fig. 16 - the original DS option, described 
in \cite{te_defl}. By reducing the iris radius $r_w \rightarrow a$ we come through Fig. 15c, Fig. 15d 
to Fig. 15e - the well known DLW. In this way an equal $e_{x0}, h_{y0}$ phasing changes into an opposite 
phasing. Transformation into another direction $r_w \rightarrow b$ leads to other structures, Fig. 15 a,
which are very compact, $2b \leq \frac{\lambda}{2}$, have a high value of $Z_e$ in SW $\pi$ mode 
and an equal $e_{x0}, h_{y0}$ phasing, similar to the initial Fig. 15b option. The deflecting field 
distribution in such structures is described by (\ref{6.1.4e}), and looks similar as the plots in 
Fig. 14 d, e. In such structures, Fig. 15a, the central bunch particle will receive, both for bunch deflection and 
for bunch rotation, a strong oscillating impact, comparable to the average deflecting field, 
and can thus be displaced to a region with strong nonlinear additions in the field. Application of these 
structures for heavy ions is tolerable due to the small transverse dimensions at low frequencies, \cite{wob}. 
Moreover, for $\beta \ll 1$ the magnetic component in the deflecting force is reduced and there is no 
substantial difference between opposite and equal $e_{x0}, h_{y0}$ phasing in (52). 
For bunch rotation at $\beta \sim 1 $, especially for multiple bunch passages, such DS's should be considered more carefully. 
With a small deterioration of the transverse emittance in each DS passage, the danger of cumulative emittance growth in the multiple 
passages appears.\\     
\begin{figure}[htb]
\centering
\epsfig{file=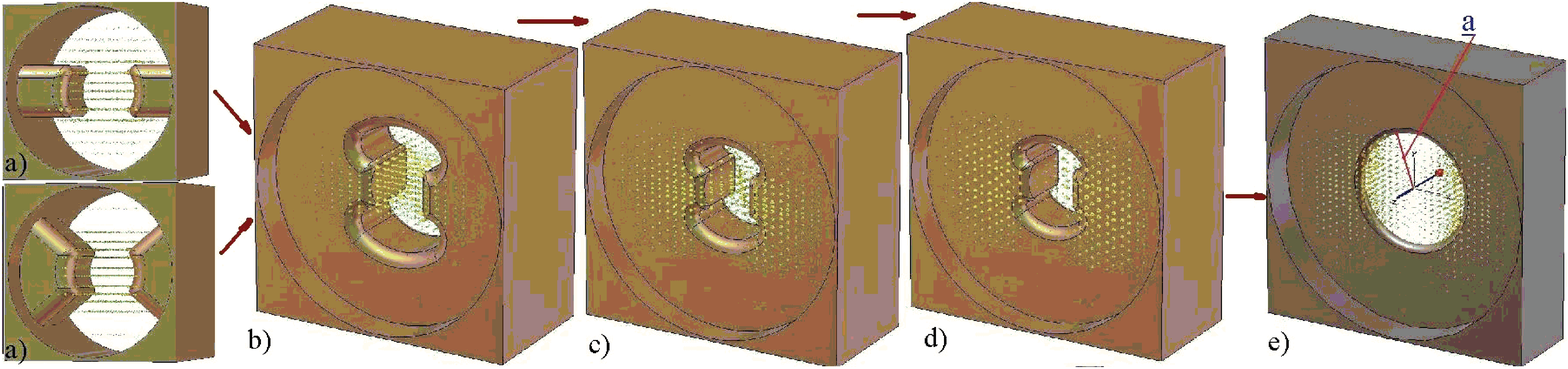, width =160.0mm}
\begin{center}
 Figure 16: Transformation of deflecting structure.
\end{center}
\label{16f}
\end{figure}
It the previous consideration we have seen the importance of the $e_{x0}, h_{y0}$ phasing 
and balance for the deflecting field quality. Most important is the phasing. For TW mode 
we can have always a tolerable field quality for low $\Theta_0$, but opposite $e_{x0}, h_{y0}$ phasing 
leads to a better $E_d$ distribution, regardless of the  $\frac{Z_0h_{y0}}{e_{x0}}$ balance. For a SW mode 
the opposite $e_{x0}, h_{y0}$ phasing also leads first to a better $E_d$ distribution and allows (second) to reduce aberrations 
in $E_d$  for the appropriate $\frac{Z_0h_{y0}}{e_{x0}}$ balance.\\
With respect to the deflecting field quality, it looks attractive to distinguish all DS in two groups - 
with opposite and with equal hybrid waves $HE_1, HM_1$ phasing. But hybrid waves $HE_1$ and $HM_1$ 
are special functions, practically inaccessible to a visual perception. We can define the phasing 
and the balance from the results of a treatment for the simulated field distribution, but it is not clear.\\
Suppose we have a DS with a pronounced predominance of the transverse electric field in the aperture  
$\vec E \approx \vec i_x E_x  \approx \vec i_r E_r$, for example the DS, shown 
in Fig. 1b, Fig. 1d or Fig. 15a. From the Maxwell equation $rot \vec E = -\frac{\partial \vec B}{\partial t}$ 
for the synchronous harmonics in (\ref{2.2e}) for each field component we find:
\begin{equation}
- i k Z_0 h_{\vartheta 0}(r,z) =\frac{\partial e_{r0}(r,z)}{\partial z} - \frac{\partial e_{z0} (r,z)}{\partial r}, 
\Rightarrow Z_0 h_{\vartheta 0}(r) = \frac{k_{z0}}{k} e_{r0}(r) - \frac{i}{k} \frac{\partial e_{z0} (r)}{\partial r}.
\label{7.1.1e}
\end{equation}
All DS with a pronounced predominance of the transverse electric field in the aperture $|E_x| \gg |E_z|$ 
have an equal phasing of the hybrid waves $HE_1$ and $HM_1$. This is valid also for DS with rods and DS utilizing a
$TM_{010}$ mode in combination with a transversely passing bunch.\\ 
To have an opposite phasing for the hybrid waves $HE_1$ and $HM_1$,
in the distribution of the original electric field the $E_z$ component should dominate, as one can see it 
in Fig. 1a, Fig. 1c and Fig. 15c.  
\section{End cell problem}
In periodical structures with a finite length the natural boundary conditions are magnetic 
or electric wall conditions at the planes of mirror symmetry. The practical case are electric 
conditions in the middle plane between adjacent disks. In this case the distributions 
for the field components in a SW cavity are perfectly the same as for a periodical structure.\\
In a real DS perfect boundary conditions are not possible at the ends of the structure -
always a beampipe is required. The field penetrating into the beampipe decays away from the 
structure but provides an initial transverse kick, which starts before the particles enter 
the actual structure. The field distribution in the beampipe is not controllable and in order
to reduce this part of the deflection and thus simultaneously reduce the total input kick the 
beampipe radius $r_t$ should be as small as reasonably possible.
\subsection{End cell for SW mode}
A model for the definition of the end cell in a SW cavity is shown in Fig. 17.
\begin{figure}[htb]
\centering
\epsfig{file=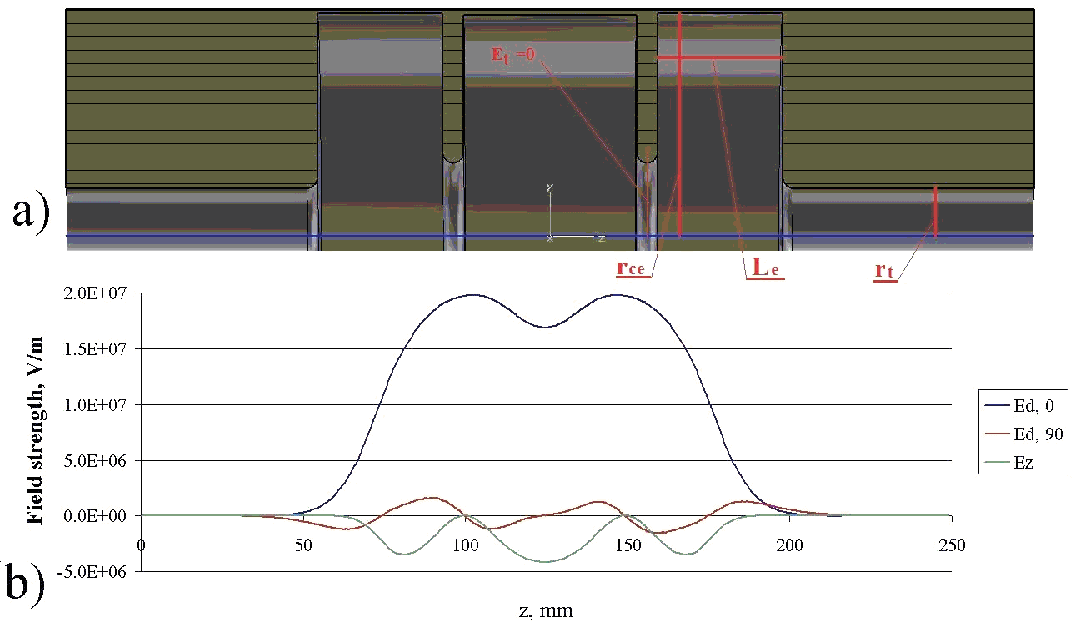, width =150.0mm}
\begin{center}
 Figure 17: Model for the definition of the end cell (a) and distributions for field components (b) -
 $E_d, \phi=0$ (blue curve), $E_d, \phi=\frac{\pi}{2}$ (red curve) and $E_z, \phi=\frac{\pi}{2}$ (green curve).
\end{center}
\label{17f}
\end{figure}
The end cells together with the beampipe are tuned to the operating frequency as separate units by 
adjusting the cell radius $r_{ce}$ while keeping the boundary condition $E_{\tau}=0$ in the middle 
of the iris connecting to the periodic structure. This ensures that the frequency and the field 
distribution are independent of the number of regular cells in the cavity.\\
By changing the length $L_e$ of the end cell the distribution of $E_d, \phi=\frac{\pi}{2}$ can be changed in a wide range,
 Fig. 18. The transverse field can be reduced but does not disappear completely.
\begin{figure}[htb]
\centering
\epsfig{file=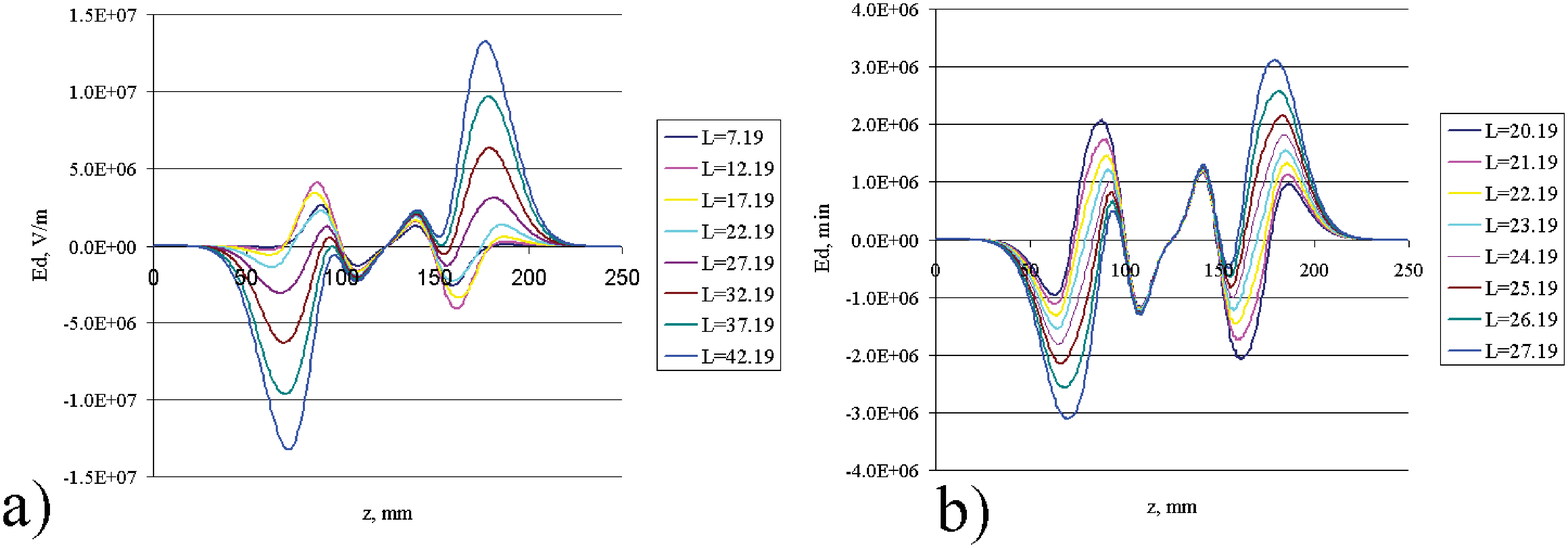, width =150.0mm}
\begin{center}
 Figure 18: Examples of input kick control in a wide range (a) and for more precise compensation 
(b) by the choice of the end cell length $L_e$.
\end{center}
\label{18f}
\end{figure}
In order to reduce the input kick the minimization of the phase deviation from the synchronous particle, following 
to (\ref{6.1.5e}) is possible. But another procedure is more convenient for a SW mode, \cite{defl_min}.\\
The first integral $Int_{1t}(z)$ 
\begin{equation}
Int_{1t}(z) =\int_{-\infty}^z E_d(z^{'},\phi=\frac{\pi}{2}) dz^{'}. 
\label{8.1.1e}
\end{equation}
is proportional to the transverse momentum and thus the transverse velocity of a particle, 
which defines the direction of the particle motion. Equivalently we can calculate the second integral $Int_{2t}(z)$, 
which is proportional to the transverse displacement of a particle. By varying the end cell length $L_e$ 
we can control either the central particle direction or displacement at a specified point.\\
The final field distributions, shown for three a cell cavity in Fig. 17b, are obtained from the 
condition $Int_{1t}(z)=0$ at the position of the first iris, i. e. the particle enters the regular 
cavity part parallel to the axis. It is realized for $L_e+\frac{t_d}{2} \approx \frac{d}{2}$ - the 
length of the end cell is approximately half of the structure period.\\
The condition $Int_{1t}(z)=0$ results in a reduced variation of $E_d,\phi=\frac{\pi}{2}$ in the end cell, 
comparable to the residual $E_d,\phi=\frac{\pi}{2}$ variation in the regular structure and keeps the central 
particle near the DS axis.
\subsection{End cell for TW mode}
The end cell problem for a TW mode requires a more difficult technique, because the end cell is 
simultaneously the RF coupler cell. Additionally to the task of reducing the input kick we have the task 
of RF matching.\\
For a TW mode the input kick in the end cell is described by phase deviations in the deflecting field 
distribution, (56) and the kick minimization is equivalent to the $|d \psi_d (z)|$ minimization in the end cell. For this 
purpose the method calculations of reflection coefficients for RF coupler matching at $\phi=0$, \cite{kroll}, 
has been extended for simultaneous $d \psi_d (z)$ determination at $\phi=\frac{\pi}{2}$.\\
\begin{figure}[htb]
\centering
\epsfig{file=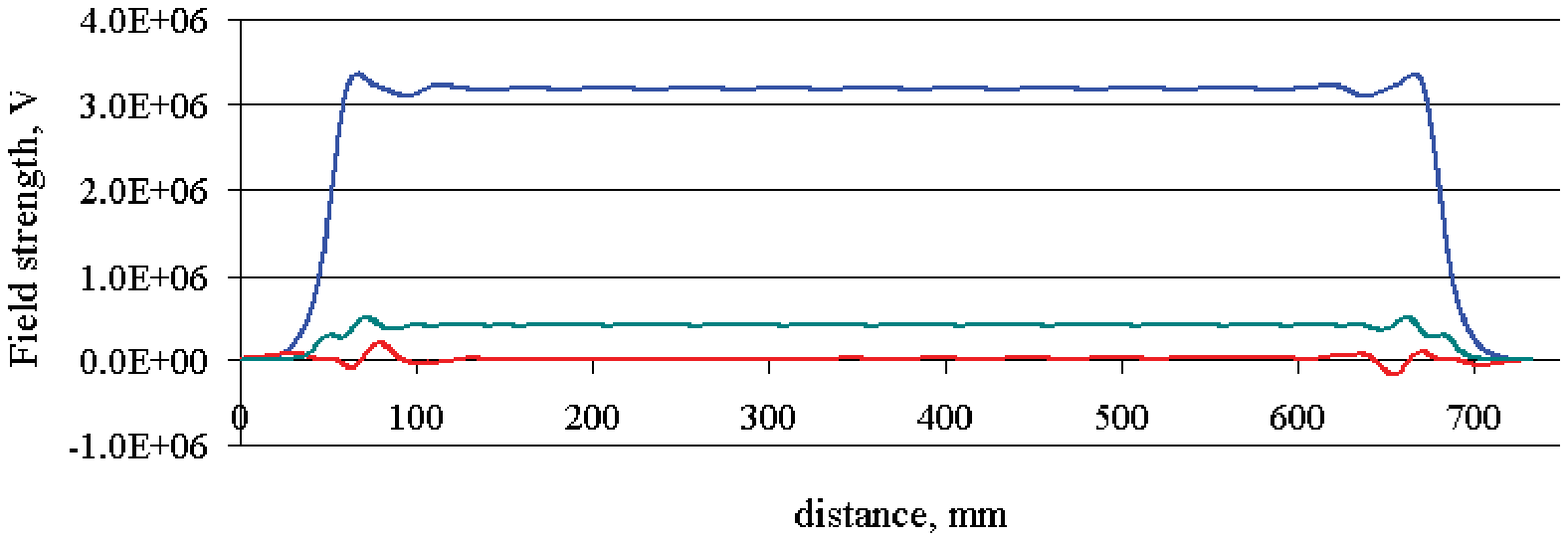, width =140.0mm}
\begin{center}
 Figure 19: Plots of $E_d$ and $E_z$ distributions for a DLW in TW mode, $\Theta_0=\frac{\pi}{3}$. 
The plots are - $E_d, \phi=0$ -  blue curve, $E_d, \phi=\frac{\pi}{2}$ -  red curve and $E_z, \phi=\frac{\pi}{2}$ - 
 green curve.
\end{center}
\label{19f}
\end{figure}
In Fig. 19 the distributions of $E_d$ and $E_z$ are plotted for a DLW in TW mode, $\Theta_0=\frac{\pi}{3}$.
Comparing with the corresponding plots in Fig. 3, one can see, at first, essentially reduced variations 
of $E_d$ and $E_z$ inside the regular structure for $\Theta_0=\frac{\pi}{3}$. It is the consequence 
of stronger harmonics attenuation for lower $\Theta_0$. But, on the background of reduced 
aberrations in the regular structure, the input kick is clearly visible.\\ 
Similar to the SW case, the input kick already starts in the beampipe. It is not a controllable region 
and the only way to reduce the kick is the reduction, if possible, of the beampipe radius.
For a large beampipe radius the value of the first integral (\ref{8.1.1e}) over the beam pipe region is essentially large and some times it is not so easy to 
compensate it with the end cell only. \\
The most essential parameter for the value of the input kick is the length of the end cell, similar to the SW case, and 
the minimal kick values are obtained for an end cell length of $\approx 0.5 d$. The plots of the field component
distributions in Fig. 19 and Fig. 5 correspond to the half cell RF couplers.\\
For the DS with visible $E_d$ deviations, Fig 5, $\Theta=\frac{2\pi}{3}$, it is not a big problem to 
reduce the input kick to the level of the $E_d$ deviations in the regular part of the DS.
If we use a DS with a more uniform field distributions, $\Theta=\frac{\pi}{3}$, Fig. 19, the input 
kick is a single source for a field nonlinearity and we see an additional effect of the RF coupler cell 
- as the deterioration of the periodicity the RF coupler perturbs the phase distribution in the adjacent DS 
cells. In Fig. 19 the maximal value of $E_d, \phi=\frac{\pi}{2}$ is displaced from the structure ends to the adjacent 
DS cells, even though the dimensions of these cells are unchanged. This effect needs more study.\\
The distribution of $E_d(z,\phi=\frac{\pi}{2})$ is always an odd function with respect to the DS middle. This means, that $Int_{1t}(z)$  and $Int_{2t}(z)$ are always even and odd, respectively, functions with 
respect to the DS middle. 
\section{Dispersion properties and limitations}
In periodical structures the main spatial harmonics provide flux propagation of the the RF power. In most 
cases the opposite $e_{r0}, h_{\vartheta 0}$ phasing at the DS axis results in a negative value of $\beta_g$. 
This is not the same statement. The opposite $e_{r0}, h_{\vartheta 0}$ phasing at the DS axis 
definitely means a part of the RF power $P_{tr} ^{-}$ in the nearest vicinity of the axis, propagates 
into negative direction. But the distribution of the field components depends on the radius $r$ which can result in a part of the RF power $P_{tr} ^{+}$ propagating into positive direction near $r \sim a$. The sign and the value 
of the group velocity depends on the total flux $P_{tr} ^{tot} = P_{tr} ^{+} + P_{tr} ^{-}$. By changing the aperture radius 
in a DLW we change the ratio $\frac{P_{tr} ^{-}}{P_{tr} ^{+}}$ and for $P_{tr} ^{-}+ P_{tr} ^{+} =0$ $\beta_g$ 
inversion is reached, (\ref{2.5.2e}). The inversion phenomenon is the consequence of the field distribution in the hybrid waves $HE_n$ and $HM_n$,
even for a single passband, when mode mixing effects are absent.\\
Together with more flexibility, the inversion phenomenon and, mainly, the dependence of the inversion point on $\Theta_0$,
 provides some limitations on the choice of the DS dimensions.\\
\begin{figure}[htb]
\centering
\epsfig{file=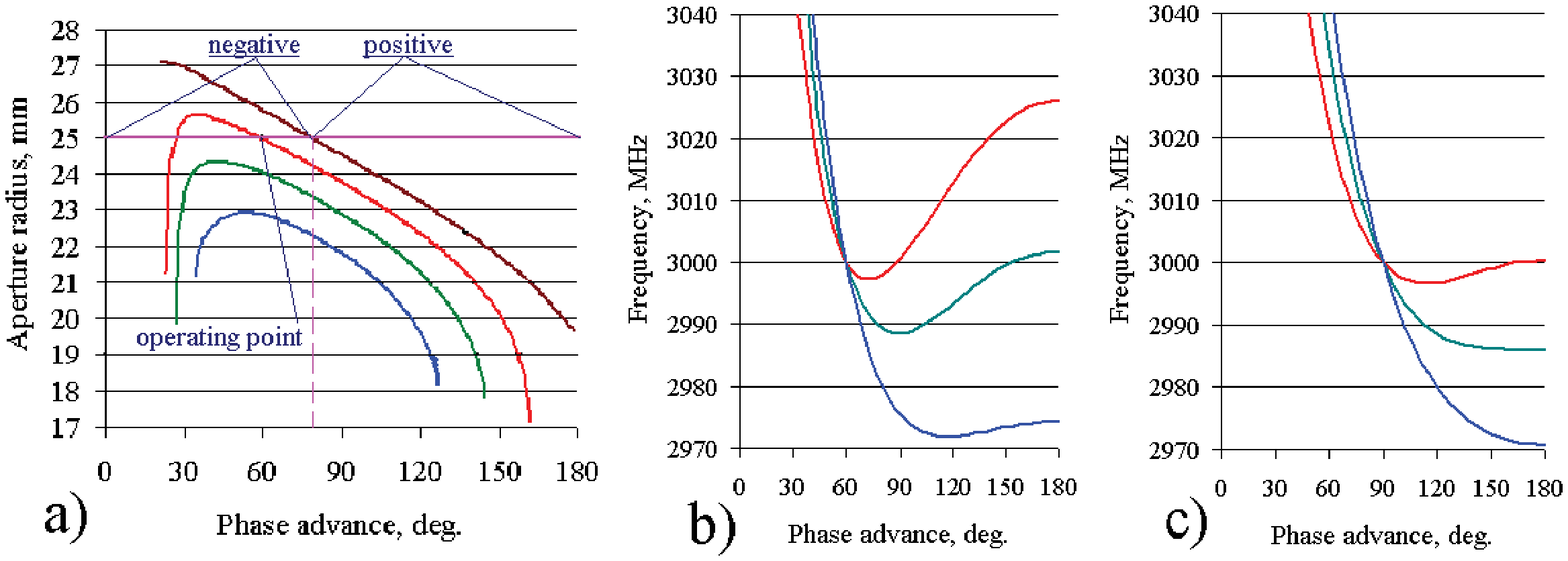, width =150.0mm}
\begin{center}
 Figure 20: The dependences of the aperture radius $a$ on $\Theta_0$ (a) and the dispersion curves for the DLW TW structure 
with $\Theta_0=\frac{\pi}{3}$ (b) and $\Theta_0=\frac{2\pi}{3}$ (c) corresponding to $\beta_g=-0.01,-0.02,-0.03$, red, green and blue 
curves, respectively.
\end{center}
\label{20f}
\end{figure} 
In Fig. 20a the plots of the dependence of aperture radius $a$ on $\Theta_0$ to get a required value of $\beta_g$ for a DLW in TW mode 
is reproduced from Fig. 8a. The inversion point dependence is plotted in Fig. 20a by a brown curve. In Fig. 20 b, c 
the dispersion curves for a DLW with different values of $\Theta_0$ and $\beta_g$ are plotted.
As one can see from Fig. 20 b, some curves have a complicated behavior - the tail of the dispersion curve 
at $\Theta \rightarrow \pi$ goes up and the curve can cross the line of the operating frequency another time.\\  
For sufficient RF efficiency the usual value is $|\beta_g| \approx (0.01 \div 0.02)$ which is not so 
far from the inversion curve. Suppose the operating point is chosen as $\Theta_0=\frac{\pi}{3}, \beta_g=-0.01$, 
as it is shown in Fig. 20a. The dispersion curve behavior can be understood  qualitatively by moving along a horizontal line $r=a_0$,  Fig. 20a. 
Staring in direction of increasing $\Theta$ we are in the region of negative 
dispersion and the mode frequency decreases. At some $\Theta$ value we cross the inversion curve and the decrease of the 
frequency comes to an end. We enter the positive dispersion region and for further increase of
$\Theta$ the mode frequency rises.\\ 
The simultaneous existence of two waves with different phase advance at the operating frequency is 
not tolerable both for RF and for beam dynamics reasons. In the classical DLW it can be avoided 
 only by limiting our choice of the operating point and restrict $\beta_g \leq -0.023$ for $\Theta_0=\frac{\pi}{3}$, 
$\beta_g \leq -0.018$ for $\Theta_0=\frac{\pi}{2}$ and $\beta_g \leq -0.01$ for $\Theta_0=\frac{2\pi}{3}$ at the expense of the
required RF power.\\    
For a SW mode the condition for the minimal value of $\Psi_{dm} \approx 2^o$ could be realized for  $\frac{B}{A} =
\frac{Z_0h_{y0}}{e_{x0}} \approx -0.87$, \cite{defl_min}, reflecting the minimal deviation of the central particle 
during bunch rotation, (\ref{6.1.4e}).\\
In general we can not derive a simple equation for the DS dispersion curve to understand the relative position 
of the point $\frac{B}{A} \approx -0.87$. According to the small pitch approximation for a DLW, from (\ref{2.5.3e}) the 
condition $\frac{B}{A} = -1.0$ corresponds to $k^2 a^2 =2$, the condition $\frac{B}{A} \approx -0.87$ corresponds 
to $k^2 a^2 \approx 2.13$ and the inversion, (\ref{2.5.2e}), corresponds to $k^2 a^2 =3$ or $\frac{B}{A} = -0.333$ 
(\ref{2.5.3e}). Results of simulations also show that a DS, which fulfills the condition of minimal $\Psi_{d}$, has a negative 
dispersion and a limited passband width $\sim (50 \div 70) MHz$, \cite{defl_min}. For the DLW SW case the dependence of the
mode frequency on the aperture radius is plotted for different $\Theta$ in Fig. 21.\\
\begin{figure}[htb]
\centering
\epsfig{file=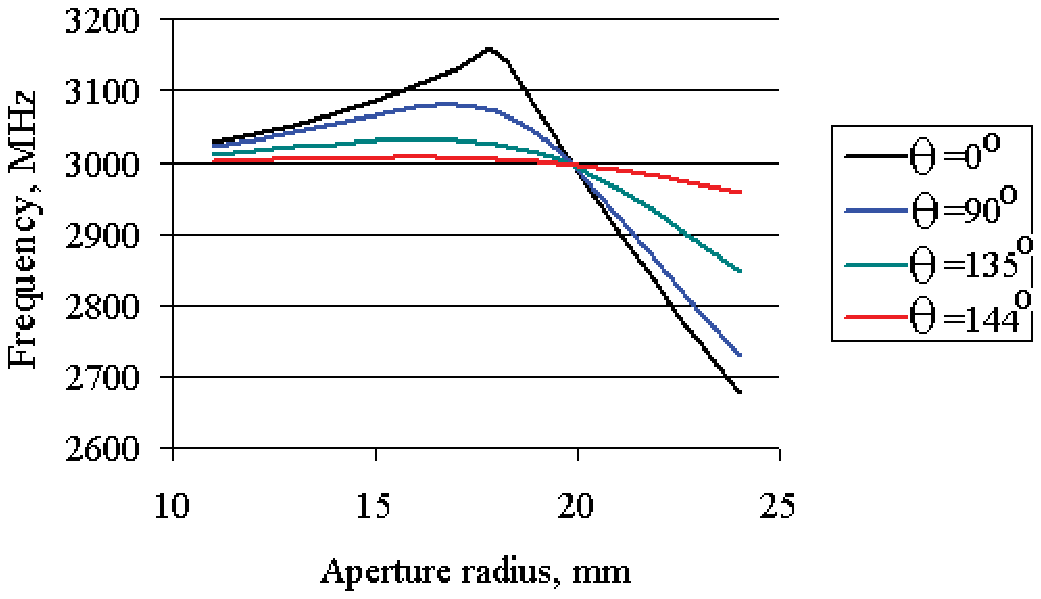, width =110.0mm}
\begin{center}
 Figure 21: Plots of the mode frequency for the DLW SW case.
\end{center}
\label{21f}
\end{figure}
Due to the inversion phenomenon, the DS dispersion curve has not a classical 'cos' - like shape, Fig. 20c, with very small 
frequency separation near the $\pi$ mode. This provides a limitation either on the possible number of cells in the SW DS with 
minimized aberrations - the minimal value of $\Psi_{dm}$, or on the possible $\Psi_{d}$ value.\\
To improve the frequency separation, the application of resonant slots is proposed in \cite{defl_min}. 
The resonant slots were also proposed for the stabilization of the deflecting plane \cite{stab}. 
One resonant slot with eigenfrequency $f_s \gg f_0$ is introduced into the disk to interact with 
the modes of the operating direction. The intensity of the slot excitation depends on both the value of $f_s$ and of
 $\Theta$ of the cavity mode. The mode frequency shift, caused by the slots,  $\delta f \sim \frac{(sin\frac{\Theta}{2})^2}{f^2_s-f^2_{\pi}}$, 
increases with $\Theta$. \\
The slots result in $E_z \neq 0 $ at the structure axis. To provide a larger $\delta f$ with smaller slot excitation 
and reduced $E_z \neq 0$ at the deflector axis, the slots in adjacent disks should be rotated by $\pi$. 
The idea of the application of slots is to increase the negative slope of dispersion curve or to decrease 
the positive slope, Fig. 20 b, for $\Theta \approx \pi$ and relax, or remove in this way the limitations 
on choice of the DS parameters from the dispersion properties. For a practical realization this idea needs 
more detailed considerations.
\section{Summary}
In the beam aperture of deflecting structure the field distribution for the dipole mode can be described by linear 
combination of hybrid waves $HE_1$ and $HM_1$. The deflecting field is composed from transverse components of electric 
and magnetic fields.\\
Only for the ultra relativistic case $\beta=1$ is the synchronous spatial harmonic in the deflecting force free from non linear additions or aberrations.
For lower particle energies even the 
synchronous harmonic has non linear additions, vanishing as $\frac{1}{\gamma^2 \beta^2}$ in the distribution.\\   
The sources of aberrations in the Lorenz force components are the higher spatial harmonics for the dipole mode 
and multipole additions. The multipole additions should be minimized in the development of the structure 
cross section. To estimate the relative level of spatial harmonics for the dipole 
mode components the criterion of maximal deviation of the field component phase from the phase of the 
synchronous harmonic is applied successfully.\\
For TW modes with low phase advance $\Theta_0 \leq \frac{\pi}{3}$ all structures have a relatively 
low level of spatial harmonics at the axis, both for the longitudinal and the transverse Lorenz force components. 
This is due to a strong attenuation of the harmonics from the aperture radius to the structure axis.\\
The level of aberrations in the $E_d$ distribution strongly depends on the phasing and the balance 
of the hybrid waves $HE_1$ and $HM_1$ in the original field. For opposite phasing the synchronous 
harmonics in the transverse components of the electric and magnetic fields work for the deflection together, but higher 
harmonics compensate. For balanced amplitudes of the hybrid waves it leads to a strongly reduced level 
of higher harmonics in the $E_d$ distribution, regardless of the level of harmonics in the original field 
components. 
This is also the only way to reduce aberrations in $E_d$ strongly for SW modes, 
$\Theta_0=\pi$, when the attenuation of harmonics in original field components is not sufficient.\\
For equal phasing of hybrid waves the level of spatial harmonics in $E_d$ distribution is enlarged 
even as compared to the level of harmonics in original field components. In the motion of  
particles enlarged oscillations with respect to the line of deflection appear. In this case the bunch as a whole can be shifted to the 
outer region with an enlarged level of nonlinear additions in the field.\\
All structures with a pronounced predominance of the transverse electric field in the aperture have an 
equal phasing of hybrid waves.\\
Due to particularities in the hybrid wave distributions exists in DS's the phenomenon 
of the group velocity inversion. In most practical cases the opposite phasing results in a negative 
dispersion and a backward wave structure. The dispersion properties lead to limitations in the 
choice of structure parameters.\\
In a DLW structure the minimization of aberrations in $E_d$ is possible at the expense of RF efficiency. 
Decoupling the control over magnetic and electric field distributions near the axis allows to combine 
both minimized aberrations and RF efficiency in the deflecting structure.
\section{Acknowledgments}
The author warmly thanks Dr. Klaus Floettmann, DESY, for support, discussion, beam dynamics expertise and for 
careful reading and correction of this paper.\\
This work was also supported in part by RBFR grant N 12 - 02 - 00654-a.


\begin{thebibliography}{99}
\addcontentsline{toc}{section}{\refname}
\bibitem{slac_sep} T.H. Fieguth, R.A. Gearhart, RF Separators and Separated Beams at SLAC,
Proc. 1975 PAC, p. 1533
\bibitem{bu_rot} R. Akre, L. Bentson, P. Emma et. al., A Transverse RF Deflecting Structure 
for Bunch Length and Phase Space Diagnostic. Proc. 2001 PAC, p. 2353
\bibitem{lapost} G. Loew, R. Neal,  Accelerating Structures, G. Dome, Review and Survey of Accelerating 
Structures, in Linear Acceletors, ed. P. Lapostolle, E. Septier, Amsterdam, 1970
\bibitem{te_defl} V. Paramonov, L. Kravchuk, S. Korepanov. Effective standing wave RF structure 
for chraged-particle beam deflector. Proc. Linac 2006 Conf., p. 469, 2006 
\bibitem{garo} Y. Garault, CERN 64-43, CERN, 1964
\bibitem{hahn} H. Hahn, Deflecting Mode in Circular Iris-Loaded Waveguides, Rev. Sci. Inst., v. 34, n. 10,
p. 1094, 1963
\bibitem{alex} J. Aleksandrov, V. Kotov, V. Vagin, P-2274, Dubna, 1975
\bibitem{pif} W. Panofsky, W. Wenzel, Some Considerations Conserning the Transverse Deflection 
of Charged Particles in Radio-Frequency Feilds. Rev. Sci. Instr., v. 27, p. 967, 1956
\bibitem{slow} R.A. Silin, V.P Sasonov, Slow wave systems, Sov. Radio, Moscow, 1966
\bibitem{mont} P. Bernard at. al., CERN 68-30, D.Alesini et. al., CTTF3-003, INFN, Frascati, 2001
\bibitem{malutin} D. Malyutin et. al., Simulations of the longitudinal 
phase space measurements with the transverse deflecting structure at PITZ, Proc. IPAC 2012, p. 637, 2012
\bibitem{comp_defl} V. Paramonov, L. Kravchuk, The compensated periodical structure for RF deflector.
Proc. Linac 2000 Conf., p.404, 2000 
\bibitem{lola} O.H. Altenmueller, R.R. Larsen, G.A. Loew, Investigations of Travelling-Wave
Separators for the Stanford Two-Mile Accelerator, The Rev. of Sci. Instr., v. 35, n 4, 1964
\bibitem{defl_min} V. Paramonov, K. Floettmann, L.V. Kravchuk, P. Orlov. Standing wave RF deflectors with 
reduced aberrations. Proc. RuPAC 2012, ISBN 978-3-954-50-125-0, p. 590, 2012 
\bibitem{kroll} N. Kroll et al., Application of time domain simulations to coupler design for periodical 
structures. Proc. Linac 2000 Conf., p. 614, 2000 
\bibitem{wob} S. Minaev et.al., Electro-Dynamics Characteristics of RF Wobbler Cell for Heavy Ion Beam. 
Proc. 2010 LINAC Conf., p. 581
\bibitem{stab} V. Paramonov, L. Kravchuk, The Resonant Method of Deflection Plane Stabilization. Proc. Linac 2010, p. 434.
\end{thebibliography}
\end{document}